\documentclass[aps,prl,showpacs,notitlepage,nofootinbib,superscriptaddress,floatfix,showkeys,twocolumn,preprintnumbers]{revtex4-1}
\usepackage{footmisc}
\usepackage{blindtext}
\usepackage{hyperref}
\usepackage{amsmath,amssymb}
\usepackage{float}
\usepackage{microtype}
\usepackage{graphicx}
\usepackage{bm}
\usepackage{latexsym}
\usepackage{epsfig}
\usepackage{psfrag}
\usepackage{color}
\usepackage[dvipsnames]{xcolor}
\usepackage{subfigure}
\usepackage{orcidlink}
\usepackage{ulem}
\usepackage[section]{placeins}
\def\e1i{\epsilon_{1\mathrm{i}}}

\allowdisplaybreaks[1]

\begin{document}
%%%%%%%%%%%%%%%%%%%%%%%%%%%%%%%%%%%%%%%%%%%%%%%%%%%%%%%%%%%%%%%%%%%%%%%%%%%%%%%
\preprint{UT-HET-140}
\preprint{RIKEN-iTHEMS-Report-24}
\preprint{KANAZAWA-24-04}
\preprint{KCL-2024-40}

\title{Supermassive black holes from inflation constrained by dark matter substructure}

\author{Shin'ichiro~Ando \orcidlink{0000-0001-6231-7693}}
\email{s.ando@uva.nl}
\affiliation{GRAPPA Institute, University of Amsterdam, 1098, XH Amsterdam, The Netherlands}
\affiliation{Kavli Institute for the Physics and Mathematics of the Universe, University of Tokyo, Chiba 277-8583, Japan}

\author{Shyam~Balaji \orcidlink{0000-0002-5364-2109}}
\email{shyam.balaji@kcl.ac.uk } 
\affiliation{Physics Department, King’s College London, Strand, London, WC2R 2LS, United Kingdom}

\author{Malcolm~Fairbairn \orcidlink{0000-0002-0566-4127}}
\email{malcolm.fairbairn@kcl.ac.uk }
\affiliation{Physics Department, King’s College London, Strand, London, WC2R 2LS, United Kingdom}
\author{Nagisa~Hiroshima \orcidlink{0000-0003-3434-0794}}
\email{hiroshima-nagisa-hd@ynu.ac.jp}
\affiliation{Department of Physics, Faculty of Engineering Science, Yokohama National University, Yokohama 240–8501, Japan}
\affiliation{Department of Physics, University of Toyama, 3190 Gofuku, Toyama 930-8555, Japan}
\affiliation{RIKEN Interdisciplinary Theoretical and Mathematical Sciences (iTHEMS),
Wako, Saitama 351-0198, Japan}
\author{Koji~Ishiwata}
\email{ishiwata@hep.s.kanazawa-u.ac.jp }
\affiliation{Institute for Theoretical Physics, Kanazawa University, Kanazawa 920-1192, Japan}

\begin{abstract}
Recent James Webb Space Telescope observations of high-redshift massive galaxy candidates have initiated renewed interest in the important mystery around the formation and evolution of our Universe's largest supermassive black holes (SMBHs). We consider the possibility that some of them were seeded by the direct collapse of primordial density perturbations from inflation into primordial black holes and analyze the consequences of this on current dark matter substructures assuming non-Gaussian primordial curvature perturbation distributions. We derive bounds on the enhanced curvature perturbation amplitude from the number of dwarf spheroidal galaxies in our Galaxy, observations of stellar streams and gravitational lensing. We find this bound region significantly overlaps with that required for SMBH seed formation and enables us to probe Gaussian and non-Gaussian curvature  perturbations corresponding to the SMBH seeds in the range ${\cal O}(10^5$\text{--}$10^{12}) M_\odot$. 
\end{abstract}
\maketitle

%%%%%%%%%%%%%%%%%%%%%%%%%%%%%%%%%%%%%%%%%%%%%%%%%%%%%%%%%%%%%%%%%%%%%%%%%%%%%%%%
\section{Introduction}
The origin of supermassive black holes (SMBHs) in our universe remains unknown.
Possible origins include stellar black holes from Population-III stars~\cite{Madau:2001sc,Bromm:2002hb}.
There are also scenarios where the direct collapse of dust clouds lead to more massive halos in which fragmentation is suppressed by some additional heating mechanism~\cite{Begelman:2006db,Volonteri:2007ax,Mayer:2007vk,Tanaka:2013boa,Izquierdo-Villalba:2023ypb}. 
 Different models for their evolution histories will be probed with future gravitational wave (GW) observations~\cite{Sesana:2010wy,Ellis:2023dgf,Ellis:2023iyb,Ellis:2024wdh}. 
As of now, it is not yet clear whether there is enough time for SMBHs to gain mass quickly enough and to do so before the early times at which they are observed in the Universe (see e.g. Ref.~\cite{Volonteri_2021} for a recent review).

The James Webb Space Telescope (JWST) is revealing the high-redshift universe in novel ways, offering much more insight into the seeds of SMBHs by studying their high redshift population. For instance, there is evidence for SMBHs at $8<z<11$ with candidates in the range $M_\textrm{BH}={\cal O}(10^6$--$10^8)M_\odot$ \cite{CEERSTeam:2023qgy,Maiolino:2023zdu,Bogdan:2023ilu,Natarajan:2023rxq,2023ApJ...959...39H}. These observations mount further pressure on models with low mass progenitors without super-Eddington accretion or rapid merger rates in the early universe.  Feedback from super-Eddington growth can hinder SMBH growth --- jets and outflows from accretion push material away, so the duration of super-Eddington growth phases should be finite~\cite{Massonneau:2022sye} but a realistic value for the duration is still unknown. 
Merger scenarios for the SMBH origin require highly clustered early populations, which can affect SMBH growth both positively and negatively, as they can act to eject SMBHs from galaxy centers \cite{Banik:2016qww}. 

These uncertainties motivate another possibility, namely that our Universe’s SMBHs did not acquire most of their mass through accretion or mergers, but are rather primordial black holes (PBHs) forming from the collapse of large density fluctuations produced during cosmic inflation. Unlike smaller PBHs, the horizon re-entry of these fluctuations should happen at later epochs, or correspondingly lower temperatures, in order to contain large-enough mass within the horizon. There are several schemes to achieve such enhanced perturbations, e.g., via a phase of ultra-slow-roll~\cite{Balaji:2022rsy,Balaji:2022dbi} or considering multiple fields \cite{Qin:2023lgo,Geller:2022nkr}.

The homogeneity and isotropy of cosmic microwave background (CMB) observations impose strict constraints on the amplitude~$A_s$ and spectral index~$n_s$ of scalar perturbations. For example, at the pivot scale $k_*=0.05~{\rm Mpc}^{-1}$, The amplitude and the spectral index is constrained as $A_s=(2.099\pm 0.029) \times 10^{-9}$ and $n_s=0.9649\pm 0.0042$ ({\it Planck} TT,TE,EE+lowE data~\cite{Planck:2018vyg}). At smaller scales, constraints on the curvature perturbations are much less stringent. Currently, the amplitude for comoving wavenumbers $k > \mathcal{O}({1})~{\rm Mpc}^{-1}$ is constrained through considerations of $\mu$- and $y$-type distortions in CMB observations~\cite{Chluba:2012we,Chluba:2015bqa}, the overproduction of PBHs~\cite{Josan:2009qn,Carr:2009jm,Byrnes:2018txb,Dalianis:2018ymb,Sato-Polito:2019hws,Gow:2020bzo}, density profiles of ultracompact minihalos~\cite{Delos:2018ueo,Nakama:2017qac}, free-free emission in the {\it Planck} foreground analysis~\cite{Abe:2021mcv}, the galaxy luminosity function~\cite{Yoshiura:2020soa} and gravitational lensing~\cite{Gilman:2021gkj}. The enhancement of the primordial power spectrum of curvature perturbations $P_\mathcal{\zeta}$ at large $k\simeq {\cal O}(10^2$--$10^4)\,\textrm{Mpc}^{-1}$ is constrained to be lower than around ${\cal O}(10^{-4})$~\footnote{The curvature perturbation in comoving coordinates $\zeta$ is used interchangeably with the comoving curvature perturbation $\mathcal{R}$ in the literature because they describe the same physical quantity during inflation and on superhorizon scales.}.

 However, the amplitude of the primordial curvature perturbation needed to explain the abundance of PBHs as SMBH seeds can be decreased if the density distribution follows non-Gaussian statistics ~\cite{Unal:2020mts,Iovino:2024uxp}. If this is the case, the current constraints listed above can be avoided.

In this work,  we highlight how the imprint of small-scale perturbations during inflation affects the evolution of hierarchical galaxy structures, as manifested in the dark matter (DM) halos and subhalos. Two quantities are considered as DM substructure probes: the number density of the dwarf spheroidal galaxies (dSphs) of the Milky Way~\cite{DES:2019vzn} and stellar stream observations~\cite{Grillmair:2006bd,Ibata:2001iv,Johnston:2001wh,Siegal-Gaskins:2007zos,Carlberg:2009ae} combined with lensing analysis. The former is representative of DM substructures with visible counterparts, while we can see the signatures of substructures that are too small to host galaxies in the latter. Previous works show that they are powerful indicators of DM properties~(e.g. Refs.~\cite{Nadler:2021dft,Banik:2019cza,Banik:2019smi,Dekker:2021scf}). We show that the same scheme is applicable for probing SMBH seeds.

The structure of this article is as follows. In Sec.~\ref{sec:pbhs}, we explain the formulation to connect primordial curvature perturbations and SMBH seed formation. Sec.~\ref{sec:dmsubstructure} is devoted to explaining the relevant physics of DM halos. We then show our main results and discuss their implications in Sec.~\ref{sec:resultsanddiscussion}, and we summarize and conclude in Sec.~\ref{sec:conclusion}.

\section{Primordial curvature perturbations and PBH abundance}
\label{sec:pbhs}
\begin{figure}[!h]
\centering
\includegraphics[width=1\linewidth]{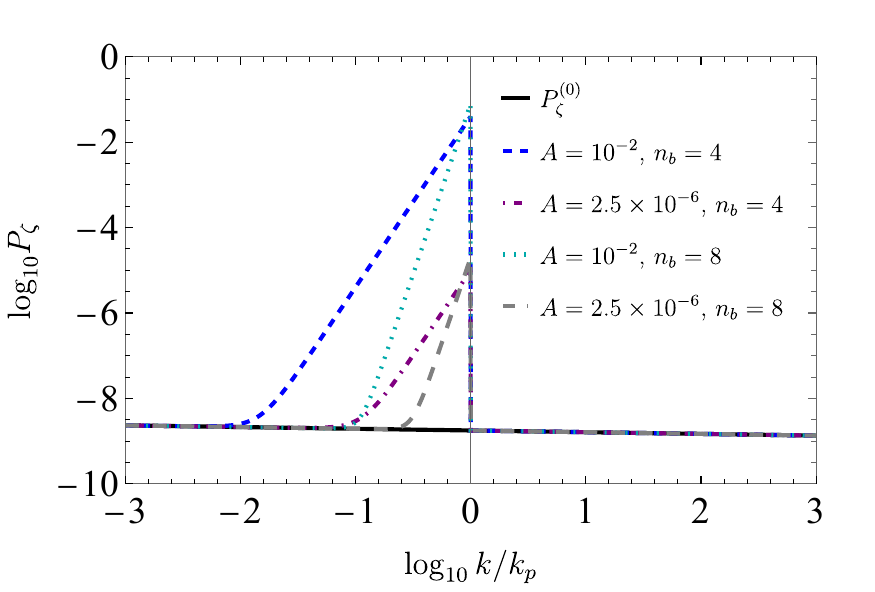}
\caption{The primordial power spectrum against the comoving-to-peak wavenumber ratio. The {\it Planck} prediction with no enhancement is shown in black-solid, while we show other combination of peak amplitude and slopes in other colors and line styles as depicted in the legend.}
\label{fig:PR}
\end{figure}

{\it Power spectrum.} 
We investigate models with a prominent feature at small scales with wavenumber $k\gtrsim \mathcal{O}({1})~{\rm Mpc}^{-1}$ in the primordial power spectrum. To evaluate the impact of the curvature perturbation in this range, we add an additional feature, namely an extra bump, on top of the nearly scale-invariant curvature perturbation spectrum that matches the features seen in the CMB

\begin{align}
 P_\zeta= P_\zeta^{(0)} +P_\zeta^{\rm bump}\,,
\end{align}
where $P_\zeta^{(0)}(k) = A_s (k/k_*)^{n_s-1}$ and
\begin{align}
  P_\zeta^{\rm bump}(k;k_p) =
    \left\{
    \begin{array}{ll}
     n_b (A-P_\zeta^{(0)}(k_p))
      \left(\frac{k}{k_p}\right)^{n_b} & k \le k_p \\
    0 & k>k_p
    \end{array}
    \right.\,.
    \label{eq:primordialpowerspectrum}
\end{align}
Here, we introduce three parameters denoted as $A$, $k_p$, and $n_b$, which describe the amplitude, the corresponding wavenumber, and the growth index of the bump, respectively.  Note the coefficent $n_b$ in Eq.~\eqref{eq:primordialpowerspectrum} is necessary to ensure that the variance of the primordial power spectrum is normalized to $A$ as per convention. Considering single-field inflation models, Ref.~\cite{Byrnes:2018txb} suggests a steepest spectral index of $n_b=4$. Conversely, Ref.~\cite{Ozsoy:2019lyy} argues that the spectral index could reach as high as $8$ after experiencing a dip in amplitude and subsequently peaking with an index less than 4. We show some illustrative examples of the primordial power spectrum with different amplitudes and growth indices as a function of the ratio $k/k_p$ in Fig. \ref{fig:PR}.

{\it Formation and evolution of PBHs.} 
We start by considering the mass of PBHs based on their time of formation in our Universe's history. See the literature (e.g. Ref.~\cite{Carr:2020gox,Escriva:2022duf} for recent reviews) for more details. PBHs that formed at later times should be much heavier, as  their sizes are governed by the size of the cosmological horizon. During the radiation-dominated era, the horizon contains the following amount of energy~\cite{Hooper:2023nnl}
\begin{align}
    M_\textrm{H}=3\times10^{9} M_\odot \left(\frac{10\,\textrm{keV}}{T}\right)^2\left(\frac{3.36}{g_*(T)}\right)^{1/2}
    \label{eq:horizonmass}
\end{align}
where $M_\textrm{H}$ is the horizon mass, $T$ is temperature of the Universe and $g_*(T)$ is the number of relativistic species at temperature $T$.  Eq.~\eqref{eq:horizonmass} provides a good estimate of the PBH mass corresponding to the temperature of the Universe at the formation epoch. When a density fluctuation is large enough, it collapses to form a PBH, the mass of which is typically smaller, given by $M_\textrm{BH} \simeq \gamma M_H$, where $\gamma\lesssim 1$ quantifies the efficiency of collapse \cite{1975ApJ...201....1C}. For example, it takes $\gamma\simeq 0.8$ for a narrow spectrum~\cite{Germani:2018jgr}. From Eq.~\eqref{eq:horizonmass}, the PBHs should have formed after ${\cal O}(1)$~MeV temperatures with horizon masses of around ${\cal O}(10^5)M_\odot$ in order to be SMBH seeds, while it should be before the onset of recombination with ${\cal O}(1)$~eV (or correspondingly, $M_H\sim{\cal O}(10^{17}) M_\odot$) corresponding to horizon masses at the incredulity limit. 

While the mass of the PBH at formation is described using the comoving wavenumber of the fluctuation~\cite{Nakama:2016gzw}, its evolution due to accretion and mergers can be parameterized by introducing two parameters ${\cal A}$ for accretion and ${\cal M}$ for mergers~\cite{Garcia-Bellido:2017aan}

\begin{align}
M_{\rm PBH, 0} &= {\cal A} {\cal M} M_{\rm seed} \nonumber \\ 
 &  \simeq  20  \gamma \cdot {\cal A} {\cal M}    \left(\frac{k}{10^6 \, {\rm Mpc}^{-1}} \right)^{-2}  M_\odot. 
\label{eqmassk}
\end{align}
Here we fix $\gamma=0.8$. Our results are not sensitive to the value of this parameter.

The initial fraction of causal horizons collapsing into PBHs can be calculated via the integral
\begin{align}
  \beta = \int_{\delta_c}^\infty P(\delta) d\delta,
\label{eqformfrac}  
\end{align}
where $P(\delta)$ is the probability distribution function of density contrast $\delta$  and $\delta_c$ is the critical density contrast required to collapse and form a black hole. The critical threshold has no universal value, and it can range anywhere between $0.3 \lesssim \delta_c \lesssim 0.66$ \cite{Sasaki:2018dmp}.

\textit{Non-Gaussianity.} The treatment of the non-Gaussianity in this work follows that provided in Ref.~\cite{Unal:2020mts} as we will summarize below. Following the literature, three types of statistics for curvature perturbations are considered: the Gaussian (G), the chi-square ($\chi^2$), and the cubic-Gaussian (G$^3$) distributions. To encompass these distributions, we express the curvature perturbation as
$\zeta = h(\zeta_{\rm G})$,
where $\zeta_{\rm G}$ is a Gaussian field. Adopting the local ansatz, the non-Gaussianity can be expressed in the following form
\begin{align}
h(\zeta_{\rm G}) = \zeta_{\rm G} + \frac{3}{5}f_{\rm NL}\left(\zeta_{\rm G}^2 - \sigma_\zeta^2\right) + \frac{9}{25}g_{\rm NL}\zeta_{\rm G}^3 + \ldots,
\label{hG}
\end{align}
where $\sigma_\zeta^2=\langle \zeta_\textrm{G}^2\rangle$ is the variance. From here, the probability distribution function for non-Gaussian $\zeta$ is calculated as
\begin{align}
P_{\rm NG}(\zeta) \mathrm{d} \zeta=\sum_{i=1}^{n}\left|\frac{\mathrm{d} h_{i}^{-1}(\zeta)}{\mathrm{d} \zeta}\right| P_{\rm G}\left(h^{-1}\right) \mathrm{d} \zeta,
\end{align}
where $h_i^{-1}(\zeta)$ is the $i$-th solution of $h(\zeta_{\rm G})=\zeta$ and $n$ is the number of the terms.
For curvature fluctuations dominantly following G-, $\chi^2$-, and G$^3$-distributions, Eq.~\eqref{hG} follows 
$h(\zeta_{\rm G})=\zeta_{\rm G}$, $h(\zeta_{\rm G})\propto (\zeta_{\rm G}^2-\sigma_\zeta^2)$, and $h(\zeta_{\rm G})\propto \zeta_{\rm G}^3$, respectively. The formation fractions are determined by evaluating the integral in Eq.~\eqref{eqformfrac}, where the density perturbations are expressed in terms of $\zeta$
~\cite{Garcia-Bellido:2016dkw,Garcia-Bellido:2017aan,Lyth:2012yp,Byrnes:2012yx}

\begin{align}
\beta_{\rm G} &= {\rm erfc} \left(  \frac{\zeta_c }{ \sqrt{2}\sigma/\sqrt{\mathcal{K}}}  \right), \nonumber\\ \nonumber\
\beta_{\chi^2} &= {\rm erfc} \left(  \sqrt{\frac{1}{2} + \frac{\zeta_c}{\sqrt{2/{\cal K}} \sigma }}  \right), \nonumber\\
\beta_{ {\rm G}^{3}} &= {\rm erfc} \left[  \left( \frac{\zeta_c }{ \sqrt{ 8  /{15\cal K}) }\sigma} \right)^{1/3} \right].
\label{eqdifferentstat}
\end{align}

We evaluate the variance $\sigma$ that enters Eq.~\eqref{eqdifferentstat} as a function of the smoothing scale $R=k^{-1}$
\begin{align}
\sigma^2(k) = \frac{16}{81}
\int d\ln k' \, \mathcal{P}_\zeta(k') \left(\frac{k'}{k}\right)^4 
\, |W(k'/k)|^2 \, T^2(k'/k)\,.
\label{eq:sigmaR}
\end{align}
Here we adopt the top–hat window function $W(x)=3(\sin x - x\cos x)/x^3$ and the radiation–era transfer function 
$T(x)=3(x/\sqrt{3})^{-3}[\sin(x/\sqrt{3})-(x/\sqrt{3})\cos(x/\sqrt{3})]$, 
as in Ref.~\cite{Kawasaki:2019mbl}. 
For a narrow peak in the primordial power spectrum, $\sigma^2(R)$ attains its maximum when 
$k_pR\simeq 2.5$–$3$, as shown in Ref.\,\cite{Kawasaki:2019mbl}.\footnote{It is pointed out in Ref.\,\cite{Ando:2018qdb} that the choice of the window function causes large uncertainties in the evaluation of the
variance. For instance, we have checked that if we use the gaussian window function, then the resultant amplitude $A$ to give $\Omega_{\rm PBH,0}^{\rm tot}=10^{-10}$, defined in Eq.\,\eqref{eq:Omega_PBH^tot}, becomes more than one order of magnitude larger.  }

When computing the variance, expressed as an integral over the external wavenumber $k$, 
we therefore evaluate the variance at the corresponding smoothing scale $R$ associated with that mode. 
This ensures that the physically relevant collapse scale is automatically incorporated by the 
window function and transfer function in Eq.~\eqref{eq:sigmaR}.

We note that $\zeta_c$ is the critical value of the curvature perturbation which represents the deviation from flatness in a region. Nonlinearities in the Press-Schechter or peak theory formalism are accounted for by introducing the factor ${\cal K}$ in the standard threshold integral. The factor $\mathcal{K}$ in the Press-Schechter formalism is set to 2 to correct for the underestimation of the mass fraction of collapsed objects by accounting for both overdense and underdense regions. This adjustment ensures that the predicted mass fraction aligns with physical expectations, effectively capturing the overall impact of non-linearities in a statistical sense. We note that mode coupling makes it hard to keep a strict $\chi^2$ distribution at small scales \cite{Young:2014oea}; but these uncertainties are negligible for our cases considering narrow spectra. For the following, we will consider when $\zeta_c=\delta_c=0.45$ \cite{Carr:1975qj,Musco:2012au}, as this is the condition required for the density contrast to form a black hole. 

\begin{figure}[!h]
\centering
\includegraphics[width=1\linewidth]{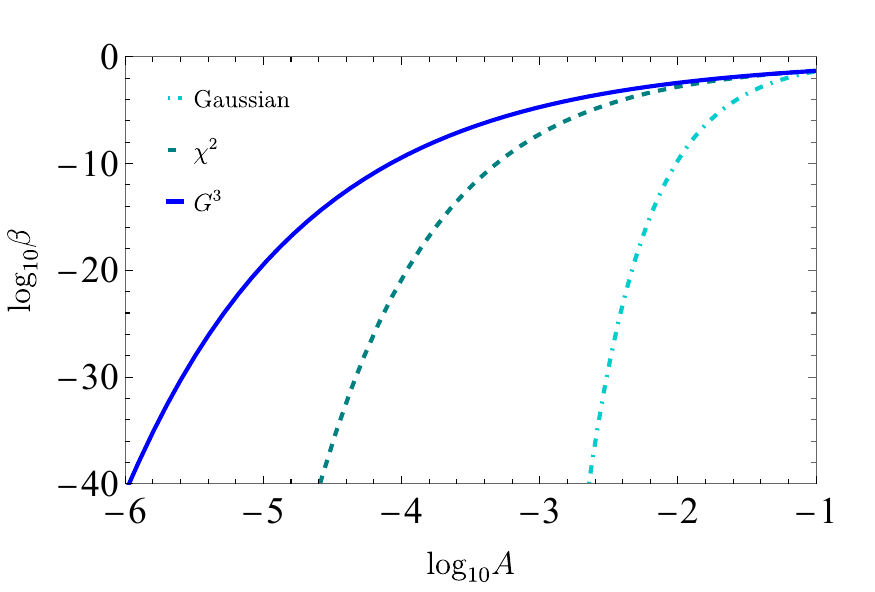}
\caption{Dependence of the initial fraction of causal horizons $\beta$ as a function of the primordial power spectrum amplitude. The Gaussian case is shown as a cyan (dot-dashed), the $\chi^2$ case as a dark cyan (dashed), and the ${\rm G}^3$ case as a blue (solid) line. The plots are corresponding to the case assuming  $\mathcal{K}=2$, $\zeta_c=\delta_c=0.45$}
\label{fig:betaPR}
\end{figure}

We show the formation fraction  in Fig.~\ref{fig:betaPR} assuming $\zeta_c=0.45$. Here we clearly see the benefits of non-Gaussianity in relaxing the power spectrum amplitude required to produce a certain fraction $\beta$ of the Universe's energy in PBHs. For example, for a power spectrum value of $P_\zeta=10^{-2}$, we get $\gtrsim10^6$ ($\gtrsim10^7$) times enhancement compared to Gaussian in the PBH fraction $\beta$ for the $\chi^2$ (G$^3$) distributions.

We relate $\beta$ to the current energy density in PBHs of seed mass $M_{\textrm{seed}}$ (described conventionally by logarithmic intervals in seed mass) as~\cite{Garcia-Bellido:2017aan,Unal:2020mts}

\begin{align}
\Omega_{\rm PBH,0} & = \frac{\rho_{\rm PBH,0}(M_{\textrm{seed}}, {\cal A}, {\cal M})}{\rho_{\rm crit,0}} \nonumber\newline \\
&\simeq2 \times 10^8 \gamma^{1/2} {\cal A} \sqrt{\frac{M_\odot }{M_\textrm{seed}}}  \beta \left( \frac{M_\textrm{seed}}{M_{\odot}}\right) .
\label{eq:pbhabundance}
\end{align}
It should be noted that the effects from accretion or mergers on the current energy density of PBHs, $\Omega_{{\rm PBH},0}$, are minor. In fact, the energy stored in the form of PBHs is not changed by mergers. Accretions would increase the energy density, however, it can be easily compensated by a change in parameter $\beta$. 
The parameter $\beta$ exhibits a high degree of sensitivity to variations in the power spectrum. An amplification by several orders of magnitude, whether through accretion or merging, can be offset by less than an order of magnitude increase in $P_\zeta$, which we will illustrate later
in Sec.~\ref{sec:resultsanddiscussion}. To calculate the total energy density of PBHs in the present Universe, we evaluate the integral
\begin{align}
    \Omega^{\rm tot}_{\rm PBH,0}&=%
     \int \Omega_{\rm PBH,0} \;  d\, \ln M_{\textrm{seed}}.
     \label{eq:Omega_PBH^tot}
\end{align}

This quantity is to be compared to the energy density of SMBHs, $\Omega_{\rm SMBH}$. Ref.~\cite{Unal:2020mts} evaluates $\Omega_{\rm SMBH}$ from the number of the galaxies that can host massive black holes~\cite{Conselice_2016} and the mass range of SMBHs as $10^6$--$10^8M_\odot$, obtaining $10^{-7}\lesssim\Omega_{\rm SMBH}\lesssim10^{-5}$ by dividing the sum of the mass in those SMBH by the total mass of the Universe. Following the estimate in the previous work, we set $\Omega^{\rm tot}_{\rm PBH,0}=10^{-10}$ as a conservative benchmark to test the PBH origin scenario for SMBHs. 

\section{Dark matter subhalos}
\label{sec:dmsubstructure}
Peaked primordial power spectra yield altered consequences for the late-time evolution of structures.
For example, models with peaked curvature perturbation spectra can be constrained with observations of DM substructures which can be traced by satellite galaxies.
Following Ref.~\cite{Ando:2022tpj}, we adopt the extended Press-Schechter (EPS) formalism to discuss the number of subhalos in a larger host halo of the Milky Way, given the aforementioned model Eq.~\eqref{eq:primordialpowerspectrum} of curvature perturbations with an enhancement parameterized by $(A, k_p, n_b)$.

In the EPS theory, instead of halo mass and redshift $(m, z)$, one adopts $(S, \delta_c^\textrm{ha})$.
$\delta_c^\textrm{ha} $ is the halo collapse threshold at redshift $z$ in the linearly extrapolated spherical collapse model, $\delta_c^\textrm{ha} \equiv 1.686/D(z)$, where $D(z)$ is the linear growth factor. $S \equiv \sigma^2(m)$ is the variance of the density fluctuation which we evaluated as that at $z=0$, where we apply
the sharp-$k$ filter. The filter is suitable for power spectra with a steep cutoff~\cite{Schneider:2013ria}. The mass fraction contained in smaller halos which collapse at redshift $z_2$, where $\delta^\textrm{ha}(z_2)$ is written $\delta^\textrm{ha}_2$, with corresponding mass scale $S_2$,
characterized by $(S_2, \delta_2^\textrm{ha})$ in their host halo $(S_1, \delta_1^\textrm{ha})$ of $z_1<z_2$ is given by
\begin{eqnarray}
    f(S_2, \delta_2^\textrm{ha}|S_1, \delta_1^\textrm{ha})dS_2 &=& \frac{1}{\sqrt{2\pi}}\frac{\delta_2^\textrm{ha}-\delta_1^\textrm{ha}}{(S_2-S_1)^{3/2}}
    \nonumber\\&&\times
    \exp\left[-\frac{(\delta_2^\textrm{ha}-\delta_1^\textrm{ha})^2}{2(S_2-S_1)}\right]dS_2.
    \nonumber\\&&
    \label{eq:EPS mass fraction}
\end{eqnarray}
With this expression, one can compute the number of subhalos that accreted on their host with masses between $m_a$ and $m_a+dm_a$ between the redshifts $z_a$ and $z_a + dz_a$ i.e. ${d^2N_{\rm sh}}/{(dm_ad z_a)}$. 
In our computation, instead of the simplest form of Eq.~(\ref{eq:EPS mass fraction}) based on the spherical collapse model, we adopt the model III 
 of Ref.~\cite{Yang:2011rf}, which better fits
 numerical simulation results  
 by adding a condition for main branch halos.

Besides $(m_a, z_a)$, the density profile of a subhalo at accretion is also characterized by its concentration parameter at the epoch,  $c_a$. Assuming the Navarro-Frenk-White (NFW) profile~\cite{Navarro:1995iw} 
\begin{equation}
    \rho(r)=\rho_s\left(\frac{r}{r_s}\right)^{-1}\left(1+\frac{r}{r_s}\right)^{-2},
\end{equation}
the concentration parameter is defined as $c_a=r_a/r_s$ where we denote the virial radius at accretion as $r_a$. The two parameters characterizing the profile, $(\rho_s,\ r_s)$ are tied to the maximum circular velocity and the corresponding radius $(V_{\rm max},\ r_{\rm max})$ as $\rho_s=(4.625/4\pi G)(V_{\rm max}/r_s)^2$, $r_s=r_{\rm max}/2.163$. 

We adopt the concentration-mass relation of Ref.~\cite{Prada:2011jf} for the following calculation. 

Once a smaller halo has accreted onto a larger one and become the subhalo, it will start losing its mass through the tidal force exerted by the host.
This process is described by the following differential equation (e.g. Ref.~\cite{vandenBosch:2004zs})
\begin{equation}
    \frac{dm}{dt} = -g \frac{m}{\tau_{\rm dyn}}\left(\frac{m}{M(z)}\right)^\eta,
\end{equation}
where $M(z)$ and $\tau_{\rm dyn}$ are the mass and dynamical time scale of the host halo, respectively, and $g$ and $\eta$ are parameters which depends on the host halo mass and redshift. 
We take values determined by fitting the tidal mass-loss rate with a single power law function~\cite{Hiroshima:2018kfv} which agrees with the numerical results~\cite{Ishiyama:2014gla}.  
 We obtain the accretion history of the host $M(z)$ (also relevant for $S_1$ in Eq.~(\ref{eq:EPS mass fraction})) through the EPS theory~\cite{Hiroshima:2022khy}, which depends on the peak parameters $(A, k_p, n_b)$ through $S_1$.
We incorporate the change of the density profile parameters $(r_s, \rho_s)$ from the evolution of $V_{\rm max}$ and $r_{\rm max}$ following the tidal mass loss of subhalos, taking the relationship derived in Ref.~\cite{Penarrubia:2010jk} which is suitable for the NFW profile, i.e., inner slope proportional to $r^{-1}$. 

\section{Results and discussion}
\label{sec:resultsanddiscussion}

In this section, we explore the constraints on curvature perturbations that can source the SMBH seeds from current observations and prospects, considering DM substructure probes of the number count of  satellite galaxies as well as stellar stream and lensing observations.  As we will see below, these show complementary performance to CMB $\mu$-type distortion measurements which were investigated in previous works~\cite{Unal:2020mts}.

\begin{figure*}
    \centering
    \includegraphics[width=8.5cm]{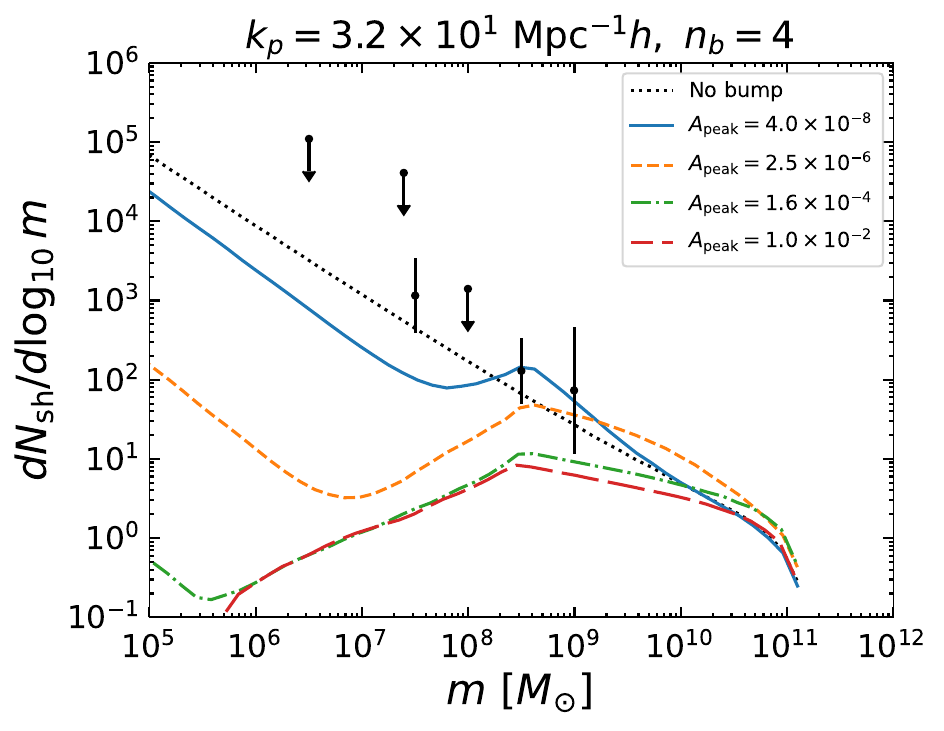}
    \includegraphics[width=8.5cm]{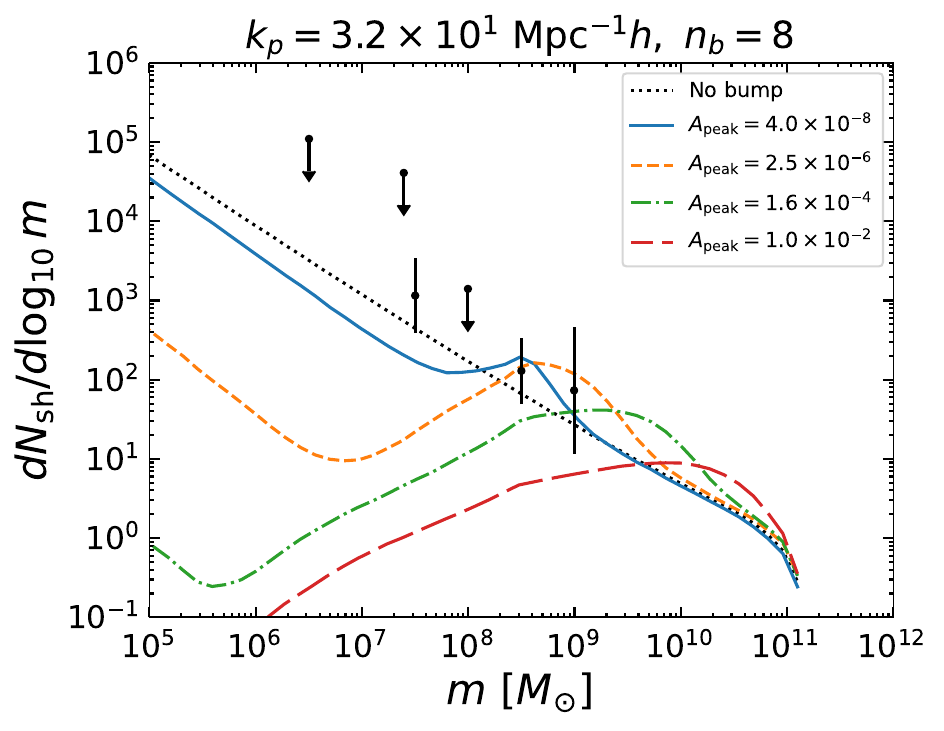}
    \caption{The subhalo mass function $dN_{\rm sh}/d \log_{10} m$ within 300~kpc volume of the Milky-Way halo. The data points are a combination of 
    the stellar-stream and lensing measurements~\cite{Banik:2019cza,Banik:2019smi}.
    The curves are predicted subhalo mass functions for $k_p = 32\,h$~Mpc$^{-1}$ and smaller to larger values of $A$ (from top to bottom). The left and right panels are for $n_b = 4$ and $n_b = 8$, respectively. The dotted line in each panel corresponds to the prediction without a bump feature. Note that the peak amplitude in the legend is simply $A_\textrm{peak}=A\, n_b$.} 
    \label{fig:SHMF}
\end{figure*}

            In Fig.~\ref{fig:SHMF}, the prediction of the subhalo mass function with several values of $A$ are plotted in different lines. The left (right) panel corresponds to the case of $n_b=4$ ($n_b=8$). We fix the parameter $k_p = 32\,h\,\mathrm{Mpc^{-1}}$ where $h=0.6736$~\cite{Planck:2018vyg}. In the calculation for $n_b=8$ case, we use the host halo mass evolution with $n_b=4$ to save computational time, since the impact is expected to be small. The figure clearly shows that the subhalo mass function within the vicinity of the Milky Way is affected by the injection of power on this characteristic scale.  
For the same wavenumber, the maximum in the subhalo mass function shifts to higher masses as the amplitude of the power enhancement increases. 
The number of smaller subhalos, on the other hand, is suppressed because halos corresponding to wavenumbers of the power spectrum bump collapse so early that smaller substructures cannot form as they will be embedded inside these larger halos.

This behavior is qualitatively consistent with numerical simulations, such as those presented in Ref.~\cite{Delos:2018ueo}. In particular, Fig.~4 of Ref. \cite{Delos:2018ueo} demonstrates that a sharp enhancement in the power spectrum results in a significant abundance of halos at that corresponding scale, while suppressing the formation of smaller structures. These trends are in good agreement with our findings based on the EPS formalism. While a detailed quantitative comparison between our semi-analytical predictions and numerical simulations remains a task for future work, preliminary comparisons with N-body simulations indicate promising agreement, supporting the validity of our model.

We also note that the choice of a filter function (e.g., sharp-$k$ or top-hat) can introduce some ambiguity when comparing different power spectrum models. However, Ref.~\cite{Ando:2022tpj} demonstrated that this ambiguity has minimal impact in cases with a sharply peaked power spectrum, supporting the reliability of our predictions.

\subsection{Constraints using satellite counts}

For a given parameter set $(A, k_p, n_b)$, we predict the number of subhalos that host dwarf satellite galaxies in them and compare this with observations. We assume that each subhalo above a certain mass hosts a galaxy. This enables us to avoid the uncertainty in galaxy formation conditions. We adopt the number of observed satellite galaxies in the Milky Way as a lower limit. Since implementing galaxy formation conditions will reduce the number of expected satellites compared with that of subhalos, our approach is conservative.

The Dark Energy Survey and PanSTARRS1 survey identified 94 satellites with the kinematic data within the virial volume of the Milky Way after imposing the completeness correction~\cite{DES:2019vzn, Dekker:2021scf}.
The satellite galaxy with the smallest velocity dispersion, $\sigma_V$, is Leo~V with $\sigma_V = 2.3$~km~s$^{-1}$~\cite{DES:2019vzn}. We take this value as the minimum for subhalos to host galaxies, and scan the parameter region where the number of subhalos satisfying the $V_{\rm max}^{\rm Leo~I} = \sqrt{3}\sigma_V = 4$~km~s$^{-1}$ is larger than 94. 

\subsection{Constraints using gravitational lensing and stellar stream data}
Stellar stream observations and gravitational lensing are pure probes of DM subhalos that induce gravitational perturbations and can provide indications to subhalos without baryonic counterparts. The existence of DM subhalos in the host will perturb the image of lensed galaxies behind the system, while
passing of DM subhalos through stellar streams will perturb the distribution of stars and creates gaps in the streams. Through these measurements, one can estimate the number of subhalos in a given mass range.

The observations of stellar streams and gravitational lensing are sensitive to the halos in the range of ${\cal O}(10^{6}$--$10^9) M_\odot$.  Each point with error bar in Fig.~\ref{fig:SHMF} denotes the subhalo mass function within 300~kpc from the Milky Way halo center, which was derived in Refs.~\cite{Banik:2019cza,Banik:2019smi}. To convert the subhalo mass function within the virial radius, as obtained from SASHIMI, to that within 300~kpc, we adopted the spatial distribution presented in Fig.~11 of Ref.~\cite{Springel:2008cc}. We interpret the points with downward arrows as limits on $N_{\rm sh}\neq 0$ due to the non-detection of stream perturbers or lensing at that mass at the 2$\sigma$ confidence level.

We performed chi-square analysis with the above data as
\begin{equation}
\label{eqchi}
    \chi^2(k_p,A,n_b) = \sum_i \frac{[N_i - N_{\rm th}(m_i|k_p,A,n_b)]^2}{\sigma_i^2},
\end{equation}
where $m_i$, $N_i$, and $\sigma_i$ are the $i$-th mass bin, mass function data value and its $1\sigma$ 
error shown in Fig.~\ref{fig:SHMF}. For simplicity, we assume that the probability distribution follows a Gaussian while it is not constrained by current observations. $N_{\rm th}(m_i|k_p,A,n_b)$ in Eq.~\eqref{eqchi} is the theoretical prediction for subhalo number at $m_i$ with the given parameters $(k_p, A, n_b)$. The parameter $n_b$ is fixed in our analysis. 
We then evaluate the excluded region on the $(k_p,A)$ plane at the 95\% confidence level by requiring 
$\Delta\chi^2(k_p,A) \equiv \chi^2(k_p,A)-\chi^2_{\rm min} > 5.99$.

In the example shown in Fig.~\ref{fig:SHMF}, we see that the model with e.g. $A n_b= 10^{-2}$ predicts too small numbers of subhalos in the mass range where lensing and stellar stream observations are sensitive, hence it can be excluded. We will also discuss future projections in the following sections.

\subsection{Sensitivity to SMBH seeds}

\begin{figure*}
\centering
\includegraphics[width=0.45\linewidth]{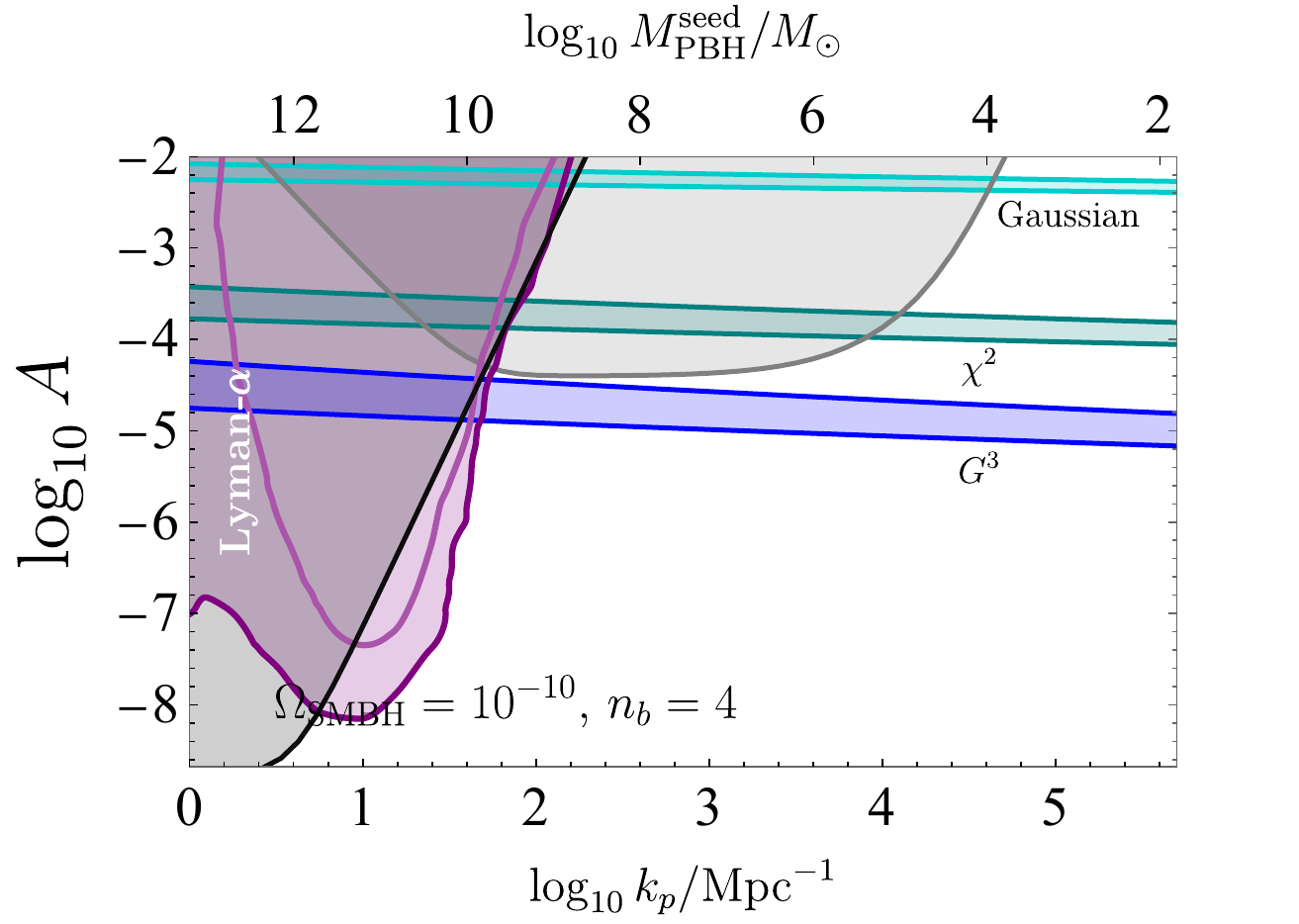}
\hskip 5pt
\includegraphics[width=0.45\linewidth]{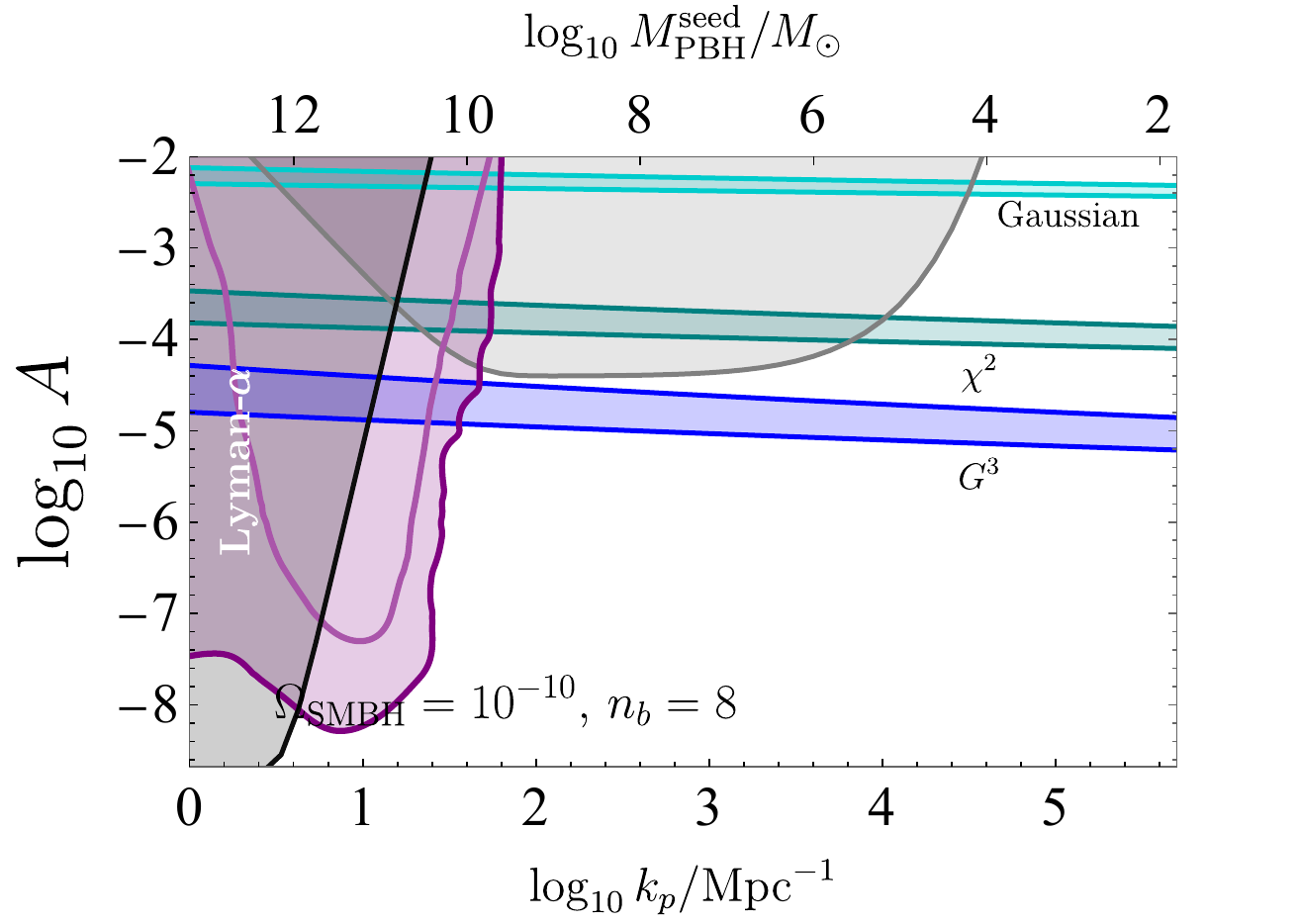}

\caption{{\bf Constraints from current observations} on the amplitude of the primordial power spectrum $A$ as a function of the peak wavenumber $k_p$ and PBH seed mass.  
The left panel assumes $n_b=4$ and the right panel does $n_b=8$. The light and dark solid purple lines represent constraints on the power spectrum from the satellite counts, and combined stream and lensing analysis at the 95\% confidence level, respectively. The solid gray line corresponds to COBE/FIRAS $\mu$-distortion bounds. The black region is excluded by Lyman-$\alpha$ data~\cite{Bird_2011}. The solid cyan, dark cyan, and blue regions correspond to the magnitude of Gaussian, $\chi^2$, and ${\rm G}^3$ distributions which would give rise to an SMBH density of $\Omega_\textrm{SMBH}=10^{-10}$, respectively.  The upper and lower boundary of these three bands correspond to the accretion factors $\mathcal{A}=1$ and $\mathcal{A}=10^5$.} 
\label{fig:sbs1}
\end{figure*}

Sensitivity to the possible SMBH seeds from current observations are summarized in Fig.~\ref{fig:sbs1}. The left (right) panel shows the case which assumes $n_b=4$ ($n_b=8$). In each panel, we plot the excluded region from satellite number counts with light purple. The constraints obtained from lensing and stellar stream observations, which are shown in dark purple, surpass those from satellite number counts. By combining the lensing and stellar stream observations, the sensitivity improves by a factor of $\simeq 10$. If the parameters $(A, k_p)$ are above these lines, our model yields fewer than 94 subhalos with $V_{\rm max} > 4$~km~s$^{-1}$, hence an upper limit at the 95\% confidence level is obtained. The sensitivity with subhalo number counts reaches as small as $A\sim 8 \times10^{-8}$ at $k_p\sim {\cal O}(10)\,{\rm Mpc}^{-1}$. 

For the Lyman-$\alpha$ exclusion shown as a dark gray region, we evaluate the full primordial power spectrum at the characteristic Lyman-$\alpha$ scale $k_{\mathrm{Ly\alpha}}\simeq2.7~{\rm Mpc^{-1}}$ and impose the empirical upper limit $P_\zeta(k_{\mathrm{Ly\alpha}})\le 3\times 10^{-9}$ derived from Lyman-$\alpha$ forest measurements~\cite{Bird_2011}.

\begin{figure*}[t!]
\centering
\includegraphics[width=0.45\linewidth]{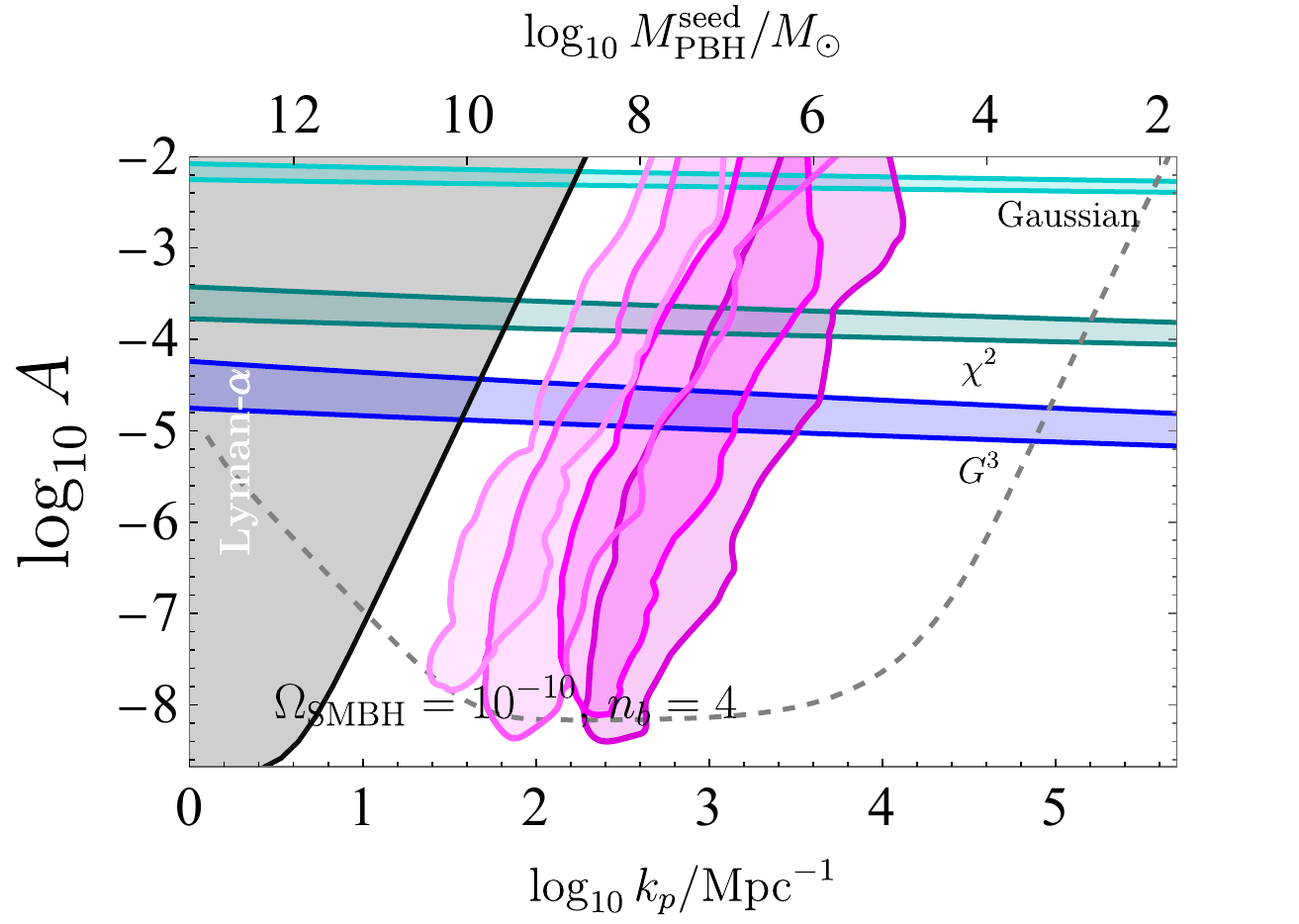}
\hskip 5pt
\includegraphics[width=0.45\linewidth]{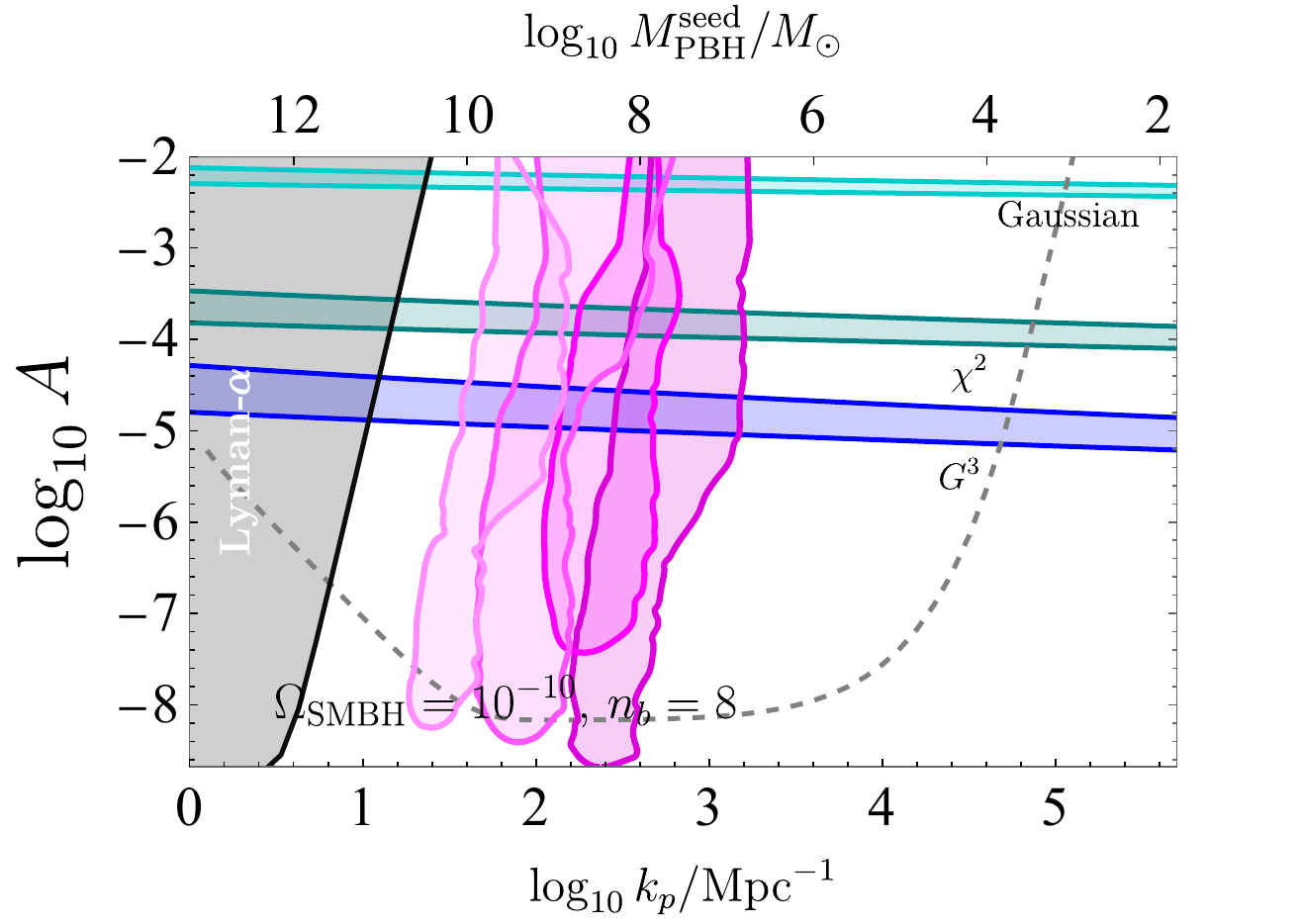}
\caption{{\bf Future prospects} on the sensitivity to the amplitude of the primordial power spectrum $A$. Projected constraints from galaxy stream observations are plotted in solid magenta lines corresponding to less than a 100\% increase in subhalos ($\frac{N_{\rm sh}(A,k_p,n_b)}{N_{\rm sh}(A=0)}< 2$) in four subhalo mass ranges, $10^5$--$10^6$, $10^6$--$10^7$, $10^7$--$10^8$, and $10^8$--$10^9 M_\odot$, denoted using the lightest to darkest magenta colours respectively. Here we assume no uncertainty in the subhalo mass function for the scale-invariant power spectrum. Prospects with future $\mu$-distortion measurements with PIXIE~\cite{Unal:2020mts} are shown with gray-dashed lines. See Fig.~\ref{fig:sbs1} for the Lyman-$\alpha$ bounds and expected amplitude for curvature perturbations to generate SMBH seeds.} 
\label{fig:sbs2}
\end{figure*}

We also plot the excluded region from CMB $\mu$-distortions in the same figure with gray-solid lines. The best current constraints on spectral distortions come from the COBE/FIRAS instrument, which finds $|\mu|\lesssim 9\times 10^{-5}$  at the 95\% confidence level \cite{Mather:1993ij,Fixsen:1996nj}. The CMB $\mu$-distortion is sensitive to the PBHs within the mass spectrum of ${\cal O}(10^4$--$10^{12}) M_\odot$. The mass range corresponds to the horizon mass associated with the enhanced primordial curvature perturbations that coincide with comoving scales entering the horizon between $z \simeq 10^6$ and recombination. The interaction of the photon-baryon fluid through the Compton scattering erases the signature of the acoustic dumping of the scalar fluctuations that enter the horizon at $z\gtrsim 10^6$. The $\mu$-distortion is calculated as~\cite{Chluba:2015bqa,Nakama:2017ohe}

\begin{align} 
\langle \mu \rangle \simeq   2.3 \int_{k_0}^\infty \frac{dk}{k} P_\zeta \left( k \right)  W(k),
\end{align}
where $P_\zeta(k)$ is convolved with a window function 

\begin{small}
\begin{align}
 W(k) =   {\rm exp} \left( - \frac{\left[ \frac{\hat k}{1360} \right]^2}{1+ \left[ \frac{\hat k}{260} \right]^{0.3} + \frac{\hat k}{340}} \right) -  {\rm exp} \left( - \left[ \frac{\hat k}{32} \right]^2 \right).
\end{align}
\end{small}

Here we denote ${\hat k} = k \,/   \, {\rm Mpc}$ and ${\hat k}_0 =1$. Current CMB $\mu$-distortion bounds are sensitive to amplitudes down to  $A \sim 4\times10^{-5}$. This probe is complementary to those DM substructures, which can be more sensitive in the mass range of $M_{\rm PBH}^{\rm seed}\gtrsim{\cal O}(10^{9})M_\odot$. 

In each panel, the amplitude of the primordial fluctuation which produces the current SMBH energy density of $\Omega_\textrm{SMBH}=10^{-10}$ assuming three different statistics, $\beta_{\rm G}, \beta_{\chi^2}$ and $\beta_{{\rm G}^3}$~\cite{Unal:2020mts} are shown with light cyan, dark cyan, and blue, respectively. Depending on the accretion efficiency ${\cal A}$, the amplitude satisfying the SMBH abundance can be in between the upper (${\cal A}=1$) or lower (${\cal A}=10^5$) boundaries of the bands. 

 The sensitivity with DM structure formation is sufficient to probe non-Gaussian statistics of $\chi^2$ and $G^3$ distributions at scales $k_p\lesssim{\cal O}(10^2){\rm Mpc}^{-1}$. For $n_b=8$, the accessible region gets slightly narrower, however, up to $k_p\sim 70\,{\rm Mpc}^{-1}$ can still be probed. These regions correspond to PBH seed masses $\gtrsim 10^{9}M_\odot$. The Gaussian statistics case over scales of $10^1\lesssim k_p \cdot \textrm{Mpc} \lesssim 3\times 10^4$ for both $n_b$ are already excluded by measurements of the CMB $\mu$-distortion by COBE/FIRAS. For a narrower range of $30\,\textrm{Mpc}^{-1}\lesssim k_p \lesssim 10^4 \textrm\,\textrm{Mpc}^{-1}$, COBE/FIRAS observations exclude the $\chi^2$ case. In all cases, the most non-Gaussian (${\rm G}^3$) case is not yet explored by $\mu$-distortion measurements.

\subsection{Prospects for future observations}

\begin{figure}[!h]
\centering
\includegraphics[width=0.93\linewidth]{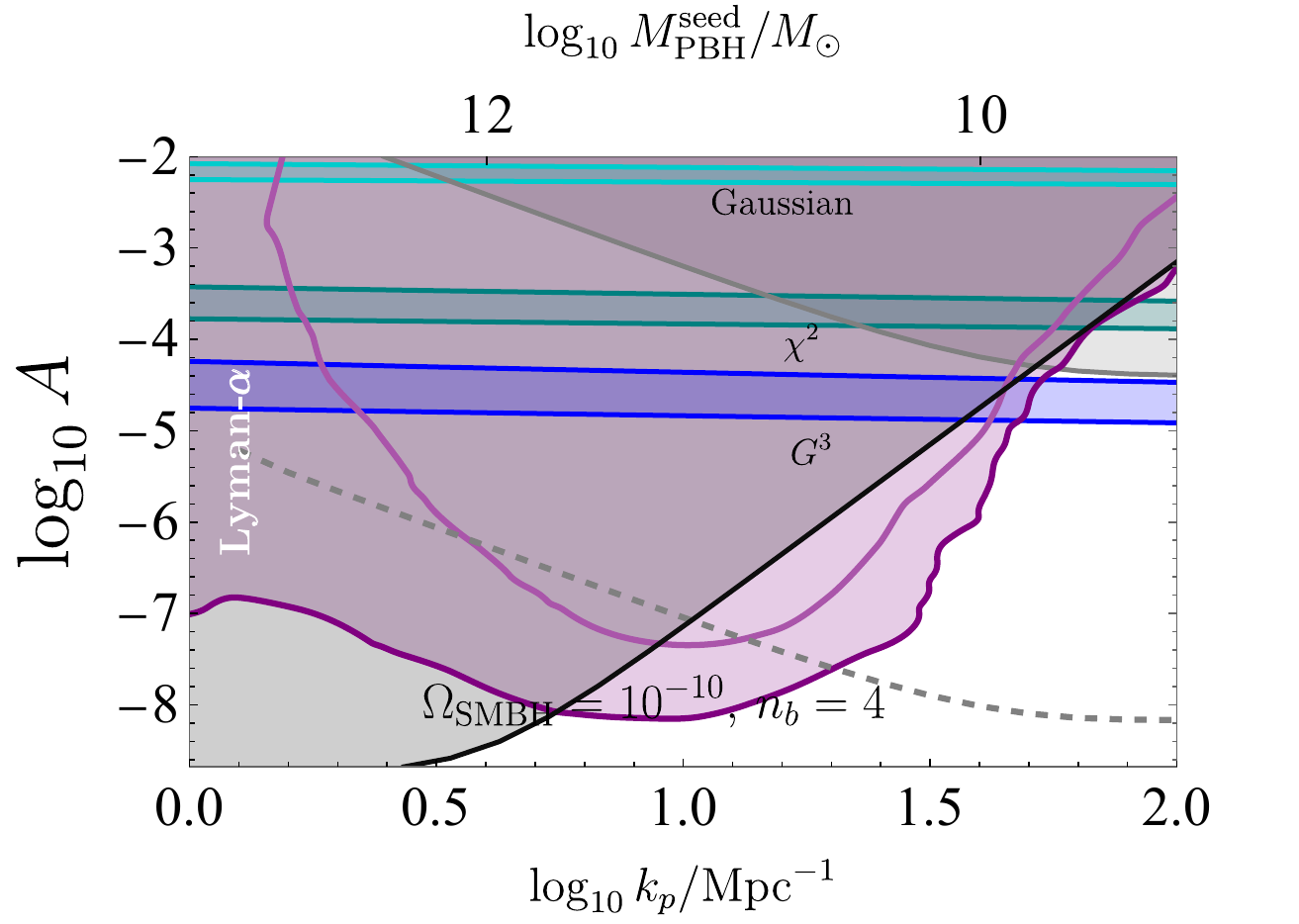}
\caption{Constraints as shown in the left panel of Fig.~\ref{fig:sbs1} with projections for the PIXIE experiment (dashed gray). From some wavenumbers the current constraints are already stronger than future projections.}
\label{fig:zoomedin}
\end{figure}

Upcoming surveys such as the Rubin Observatory Legacy Survey of Space and Time (LSST)~\cite{S5FIRE:2022ylq} can extend these constraints to much smaller scales. In Fig.~\ref{fig:sbs2}, we showcase examples assuming ${N_{\rm sh}(A,k_p,n_b)}/{N_{\rm sh}(A=0)}< 2$ for four subhalo mass ranges, $10^5$--$10^6$, $10^6$--$10^7$, $10^7$--$10^8$, and $10^8$--$10^9 M_\odot$ with light to dark magenta. Here we parametrize the abundance ratio by taking the ratio of the abundance to the case with $A=0$, i.e. without an enhancement in the power spectrum. 
Using this approach, we highlight that there is strong potential to explore the PBH origin of the SMBH by probing its mass down to $\simeq {\cal O}(10^5) M_\odot$ for $n_b=4$ and $\simeq {\cal O}(10^7)M_\odot$ for $n_b=8$.  SMBH seeds of mass $10^5$--$10^6 M_\odot$ would naturally explain the observed black holes at high redshifts by JWST. In this approach, we can probe power spectrum amplitudes down to $\simeq {\cal O}(10^{-8})$ for a certain range of scales. The constraints can be comparably strong to those expected with future $\mu$-distortion measurements by PIXIE~\cite{Unal:2020mts}, which are shown with dashed lines in the figure. Also for $n_b=8$, the DM substructure probe can be advantageous relative to to $\mu$-distortion measurements. We also remark in Fig.~\ref{fig:zoomedin} the relative strength of the current DM substructure constraints for small $k_p$ to future $\mu$-distortion probes with PIXIE.

%%%%%%%%%%%%%%%%%%%%%%%%%%%%%%%%%%%%%%%%%%%%%%%%%%%%%%%%%%%%%%%%%%%%%%%%%%%%%%%%
\section{Discussion and conclusions}
\label{sec:conclusion}

In this work, we have studied features in the primordial curvature perturbation power spectrum that give rise to over-densities which collapse to form PBHs large enough to serve as progenitors for SMBHs. 
As previously discussed in Ref.~\cite{Unal:2020mts}, PBH progenitor scenarios for SMBH seeds are in conflict with CMB $\mu$-distortion observations if a Gaussian distribution is assumed for the primordial curvature perturbations. If primordial curvature perturbations follow non-Gaussian distributions, such constraints can be evaded hence increased sensitivities from future observations are required to test such models.

In such situations, the mass function of subhalos around the Milky Way would also be altered, leading to different numbers of satellite galaxies and smaller dark subhalos. We investigated the region of parameter space which can be constrained from current observations on the number counts of the satellite galaxies and dark subhalo estimates with stellar stream and lensing observations. The DM substructure analysis excludes large regions of amplitude for peak wavenumbers less than ${\cal O}(100)$ Mpc$^{-1}$, corresponding to PBH seed masses greater than $\simeq {\cal O}(10^9) M_\odot$, which cannot be probed with current CMB $\mu$-distortion measurements.

For models with a peaked power spectrum, subhalos corresponding to the peak mass scale are expected to have significantly higher concentrations. These enhanced densities can influence a variety of astrophysical observables including the ones explored in this work. While our satellite constraints indirectly include such structural information as the models are constrained by the number of satellites with $V_{\rm max} > 4 ~ \mathrm{km~s^{-1}}$, we primarily focused on the subhalo mass function. We acknowledge that a full analysis incorporating the internal density structure of subhalos would provide additional insights (e.g., \cite{Esteban:2023xpk}).

Potential for future observations to extend the probable range with the scheme of this article is also discussed. We highlight the possibility of probing the PBH origin of SMBHs down to masses of approximately ${\cal O}(10^5) M_\odot$ and power spectrum amplitudes as low as ${\cal O}(10^{-7})$ over certain scales. We have shown that DM substructure bounds are comparably powerful to future CMB $\mu$-distortion measurements for probing the enhanced curvature perturbations. Recent works indicate that the second-order gravitational waves (e.g. Ref.~\cite{Byrnes:2018txb}) and Lyman-$\alpha$ data (e.g. Ref.~\cite{ragavendra2024}) could also be important for probing certain models which predict enhanced curvature perturbations. Those probes may have comparable sensitivity in some regions to our work, however, we remark that the prospects shown here with DM substructures are conservatively obtained. There has recently been an effort to propagate non-Gaussian effects onto the $\mu$-distortion bound in Refs.~\cite{Sharma:2024img,Byrnes:2024vjt}. However, the effects found in those works are significant only at wavenumbers $k_p\gtrsim 10^4$-$10^5$ Mpc$^{-1}$, which is complementary to our work. It should be noted though that the limits derived in those works are not directly comparable to ours because of the different templates used for the power spectrum. PBHs and DM subhalos originate from the same primordial power spectrum, hence both of their respective formations are influenced by the shape of the probability distribution function of these fluctuations. Non-Gaussianity that thickens the tails of the probability distribution function (as studied in this work) where large fluctuations occur, would result in a larger number of regions being dense enough to collapse into DM subhalos. This would further strengthen our bounds if we included such effects in our analysis. We intend to study the effect of non-Gaussianity on DM subhalos in detail in a future work while presenting the most conservative scenario in this one.

Note: While this work was under revision, Ref.~\cite{Nadler:2025crd} presented numerical simulations of subhalo abundances for inflationary spectra containing moderate log-normal bumps at wavenumbers $k\simeq30~{\rm Mpc^{-1}}$.  They reported no significant suppression of low-mass halos, which is consistent with our predictions (see Fig.~\ref{fig:SHMF}), where the subhalo mass function remains essentially unmodified for comparable amplitudes and scales.  Their results therefore complement our semi-analytical analysis by confirming the expected behavior in the weak power spectrum enhancement regime.

\section{Acknowledgements}
We would like to thank Antonio Iovino, Gabriele Franciolini, Christian Byrnes and Devanshu Sharma for useful feedback. We are grateful to Masahiro Kawasaki for valuable and insightful discussions. SB and MF are supported by the STFC under grant ST/X000753/1. SA and MF are extremely grateful for support via the Royal Society International Exchange project ``Probing particle nature of dark matter using small-scale distribution''  which made this work possible. The work of SA was also supported by MEXT KAKENHI under grant numbers JP20H05850, JP20H05861, and JP24K07039.
The work of NH was partly supported by MEXT KAKENHI Grant Number 20H05852, 22K14035, and MEXT Leading Initiative for Excellent Young Researchers Grant Number 2023L0013.  The work of KI was supported by JSPS KAKENHI Grant Number JP20H01894, JP25K07317,  and JSPS Core-to-Core Program Grant No. JPJSCCA20200002.

%%%%%%%%%%%%%%%%%%%%%%%%%%%%%%%%%%%%%%%%%%%%%%%%%%%%%%%%%%%%%%%%%%%%%%%%%%%%%%%%
\bibliographystyle{apsrev4-1}
\bibliography{references.bib}

\begin{thebibliography}{88}%
\makeatletter
\providecommand \@ifxundefined [1]{%
 \@ifx{#1\undefined}
}%
\providecommand \@ifnum [1]{%
 \ifnum #1\expandafter \@firstoftwo
 \else \expandafter \@secondoftwo
 \fi
}%
\providecommand \@ifx [1]{%
 \ifx #1\expandafter \@firstoftwo
 \else \expandafter \@secondoftwo
 \fi
}%
\providecommand \natexlab [1]{#1}%
\providecommand \enquote  [1]{``#1''}%
\providecommand \bibnamefont  [1]{#1}%
\providecommand \bibfnamefont [1]{#1}%
\providecommand \citenamefont [1]{#1}%
\providecommand \href@noop [0]{\@secondoftwo}%
\providecommand \href [0]{\begingroup \@sanitize@url \@href}%
\providecommand \@href[1]{\@@startlink{#1}\@@href}%
\providecommand \@@href[1]{\endgroup#1\@@endlink}%
\providecommand \@sanitize@url [0]{\catcode `\\12\catcode `\$12\catcode `\&12\catcode `\#12\catcode `\^12\catcode `\_12\catcode `\%12\relax}%
\providecommand \@@startlink[1]{}%
\providecommand \@@endlink[0]{}%
\providecommand \url  [0]{\begingroup\@sanitize@url \@url }%
\providecommand \@url [1]{\endgroup\@href {#1}{\urlprefix }}%
\providecommand \urlprefix  [0]{URL }%
\providecommand \Eprint [0]{\href }%
\providecommand \doibase [0]{http://dx.doi.org/}%
\providecommand \selectlanguage [0]{\@gobble}%
\providecommand \bibinfo  [0]{\@secondoftwo}%
\providecommand \bibfield  [0]{\@secondoftwo}%
\providecommand \translation [1]{[#1]}%
\providecommand \BibitemOpen [0]{}%
\providecommand \bibitemStop [0]{}%
\providecommand \bibitemNoStop [0]{.\EOS\space}%
\providecommand \EOS [0]{\spacefactor3000\relax}%
\providecommand \BibitemShut  [1]{\csname bibitem#1\endcsname}%
\let\auto@bib@innerbib\@empty
%</preamble>
\bibitem [{\citenamefont {Madau}\ and\ \citenamefont {Rees}(2001)}]{Madau:2001sc}%
  \BibitemOpen
  \bibfield  {author} {\bibinfo {author} {\bibfnamefont {P.}~\bibnamefont {Madau}}\ and\ \bibinfo {author} {\bibfnamefont {M.~J.}\ \bibnamefont {Rees}},\ }\href {\doibase 10.1086/319848} {\bibfield  {journal} {\bibinfo  {journal} {Astrophys. J. Lett.}\ }\textbf {\bibinfo {volume} {551}},\ \bibinfo {pages} {L27} (\bibinfo {year} {2001})},\ \Eprint {http://arxiv.org/abs/astro-ph/0101223} {arXiv:astro-ph/0101223} \BibitemShut {NoStop}%
\bibitem [{\citenamefont {Bromm}\ and\ \citenamefont {Loeb}(2003)}]{Bromm:2002hb}%
  \BibitemOpen
  \bibfield  {author} {\bibinfo {author} {\bibfnamefont {V.}~\bibnamefont {Bromm}}\ and\ \bibinfo {author} {\bibfnamefont {A.}~\bibnamefont {Loeb}},\ }\href {\doibase 10.1086/377529} {\bibfield  {journal} {\bibinfo  {journal} {Astrophys. J.}\ }\textbf {\bibinfo {volume} {596}},\ \bibinfo {pages} {34} (\bibinfo {year} {2003})},\ \Eprint {http://arxiv.org/abs/astro-ph/0212400} {arXiv:astro-ph/0212400} \BibitemShut {NoStop}%
\bibitem [{\citenamefont {Begelman}\ \emph {et~al.}(2006)\citenamefont {Begelman}, \citenamefont {Volonteri},\ and\ \citenamefont {Rees}}]{Begelman:2006db}%
  \BibitemOpen
  \bibfield  {author} {\bibinfo {author} {\bibfnamefont {M.~C.}\ \bibnamefont {Begelman}}, \bibinfo {author} {\bibfnamefont {M.}~\bibnamefont {Volonteri}}, \ and\ \bibinfo {author} {\bibfnamefont {M.~J.}\ \bibnamefont {Rees}},\ }\href {\doibase 10.1111/j.1365-2966.2006.10467.x} {\bibfield  {journal} {\bibinfo  {journal} {Mon. Not. Roy. Astron. Soc.}\ }\textbf {\bibinfo {volume} {370}},\ \bibinfo {pages} {289} (\bibinfo {year} {2006})},\ \Eprint {http://arxiv.org/abs/astro-ph/0602363} {arXiv:astro-ph/0602363} \BibitemShut {NoStop}%
\bibitem [{\citenamefont {Volonteri}\ \emph {et~al.}(2008)\citenamefont {Volonteri}, \citenamefont {Lodato},\ and\ \citenamefont {Natarajan}}]{Volonteri:2007ax}%
  \BibitemOpen
  \bibfield  {author} {\bibinfo {author} {\bibfnamefont {M.}~\bibnamefont {Volonteri}}, \bibinfo {author} {\bibfnamefont {G.}~\bibnamefont {Lodato}}, \ and\ \bibinfo {author} {\bibfnamefont {P.}~\bibnamefont {Natarajan}},\ }\href {\doibase 10.1111/j.1365-2966.2007.12589.x} {\bibfield  {journal} {\bibinfo  {journal} {Mon. Not. Roy. Astron. Soc.}\ }\textbf {\bibinfo {volume} {383}},\ \bibinfo {pages} {1079} (\bibinfo {year} {2008})},\ \Eprint {http://arxiv.org/abs/0709.0529} {arXiv:0709.0529 [astro-ph]} \BibitemShut {NoStop}%
\bibitem [{\citenamefont {Mayer}\ \emph {et~al.}(2007)\citenamefont {Mayer}, \citenamefont {Kazantzidis}, \citenamefont {Madau}, \citenamefont {Colpi}, \citenamefont {Quinn},\ and\ \citenamefont {Wadsley}}]{Mayer:2007vk}%
  \BibitemOpen
  \bibfield  {author} {\bibinfo {author} {\bibfnamefont {L.}~\bibnamefont {Mayer}}, \bibinfo {author} {\bibfnamefont {S.}~\bibnamefont {Kazantzidis}}, \bibinfo {author} {\bibfnamefont {P.}~\bibnamefont {Madau}}, \bibinfo {author} {\bibfnamefont {M.}~\bibnamefont {Colpi}}, \bibinfo {author} {\bibfnamefont {T.~R.}\ \bibnamefont {Quinn}}, \ and\ \bibinfo {author} {\bibfnamefont {J.}~\bibnamefont {Wadsley}},\ }\href {\doibase 10.1126/science.1141858} {\bibfield  {journal} {\bibinfo  {journal} {Science}\ }\textbf {\bibinfo {volume} {316}},\ \bibinfo {pages} {1874} (\bibinfo {year} {2007})},\ \Eprint {http://arxiv.org/abs/0706.1562} {arXiv:0706.1562 [astro-ph]} \BibitemShut {NoStop}%
\bibitem [{\citenamefont {Tanaka}\ and\ \citenamefont {Li}(2014)}]{Tanaka:2013boa}%
  \BibitemOpen
  \bibfield  {author} {\bibinfo {author} {\bibfnamefont {T.~L.}\ \bibnamefont {Tanaka}}\ and\ \bibinfo {author} {\bibfnamefont {M.}~\bibnamefont {Li}},\ }\href {\doibase 10.1093/mnras/stu042} {\bibfield  {journal} {\bibinfo  {journal} {Mon. Not. Roy. Astron. Soc.}\ }\textbf {\bibinfo {volume} {439}},\ \bibinfo {pages} {1092} (\bibinfo {year} {2014})},\ \Eprint {http://arxiv.org/abs/1310.0859} {arXiv:1310.0859 [astro-ph.CO]} \BibitemShut {NoStop}%
\bibitem [{\citenamefont {Izquierdo-Villalba}\ \emph {et~al.}(2023)\citenamefont {Izquierdo-Villalba}, \citenamefont {Colpi}, \citenamefont {Volonteri}, \citenamefont {Spinoso}, \citenamefont {Bonoli},\ and\ \citenamefont {Sesana}}]{Izquierdo-Villalba:2023ypb}%
  \BibitemOpen
  \bibfield  {author} {\bibinfo {author} {\bibfnamefont {D.}~\bibnamefont {Izquierdo-Villalba}}, \bibinfo {author} {\bibfnamefont {M.}~\bibnamefont {Colpi}}, \bibinfo {author} {\bibfnamefont {M.}~\bibnamefont {Volonteri}}, \bibinfo {author} {\bibfnamefont {D.}~\bibnamefont {Spinoso}}, \bibinfo {author} {\bibfnamefont {S.}~\bibnamefont {Bonoli}}, \ and\ \bibinfo {author} {\bibfnamefont {A.}~\bibnamefont {Sesana}},\ }\href {\doibase 10.1051/0004-6361/202347008} {\bibfield  {journal} {\bibinfo  {journal} {Astron. Astrophys.}\ }\textbf {\bibinfo {volume} {677}},\ \bibinfo {pages} {A123} (\bibinfo {year} {2023})},\ \Eprint {http://arxiv.org/abs/2305.16410} {arXiv:2305.16410 [astro-ph.GA]} \BibitemShut {NoStop}%
\bibitem [{\citenamefont {Sesana}\ \emph {et~al.}(2011)\citenamefont {Sesana}, \citenamefont {Gair}, \citenamefont {Berti},\ and\ \citenamefont {Volonteri}}]{Sesana:2010wy}%
  \BibitemOpen
  \bibfield  {author} {\bibinfo {author} {\bibfnamefont {A.}~\bibnamefont {Sesana}}, \bibinfo {author} {\bibfnamefont {J.}~\bibnamefont {Gair}}, \bibinfo {author} {\bibfnamefont {E.}~\bibnamefont {Berti}}, \ and\ \bibinfo {author} {\bibfnamefont {M.}~\bibnamefont {Volonteri}},\ }\href {\doibase 10.1103/PhysRevD.83.044036} {\bibfield  {journal} {\bibinfo  {journal} {Phys. Rev. D}\ }\textbf {\bibinfo {volume} {83}},\ \bibinfo {pages} {044036} (\bibinfo {year} {2011})},\ \Eprint {http://arxiv.org/abs/1011.5893} {arXiv:1011.5893 [astro-ph.CO]} \BibitemShut {NoStop}%
\bibitem [{\citenamefont {Ellis}\ \emph {et~al.}(2024{\natexlab{a}})\citenamefont {Ellis}, \citenamefont {Fairbairn}, \citenamefont {H\"utsi}, \citenamefont {Raidal}, \citenamefont {Urrutia}, \citenamefont {Vaskonen},\ and\ \citenamefont {Veerm\"ae}}]{Ellis:2023dgf}%
  \BibitemOpen
  \bibfield  {author} {\bibinfo {author} {\bibfnamefont {J.}~\bibnamefont {Ellis}}, \bibinfo {author} {\bibfnamefont {M.}~\bibnamefont {Fairbairn}}, \bibinfo {author} {\bibfnamefont {G.}~\bibnamefont {H\"utsi}}, \bibinfo {author} {\bibfnamefont {J.}~\bibnamefont {Raidal}}, \bibinfo {author} {\bibfnamefont {J.}~\bibnamefont {Urrutia}}, \bibinfo {author} {\bibfnamefont {V.}~\bibnamefont {Vaskonen}}, \ and\ \bibinfo {author} {\bibfnamefont {H.}~\bibnamefont {Veerm\"ae}},\ }\href {\doibase 10.1103/PhysRevD.109.L021302} {\bibfield  {journal} {\bibinfo  {journal} {Phys. Rev. D}\ }\textbf {\bibinfo {volume} {109}},\ \bibinfo {pages} {L021302} (\bibinfo {year} {2024}{\natexlab{a}})},\ \Eprint {http://arxiv.org/abs/2306.17021} {arXiv:2306.17021 [astro-ph.CO]} \BibitemShut {NoStop}%
\bibitem [{\citenamefont {Ellis}\ \emph {et~al.}(2023)\citenamefont {Ellis}, \citenamefont {Fairbairn}, \citenamefont {Urrutia},\ and\ \citenamefont {Vaskonen}}]{Ellis:2023iyb}%
  \BibitemOpen
  \bibfield  {author} {\bibinfo {author} {\bibfnamefont {J.}~\bibnamefont {Ellis}}, \bibinfo {author} {\bibfnamefont {M.}~\bibnamefont {Fairbairn}}, \bibinfo {author} {\bibfnamefont {J.}~\bibnamefont {Urrutia}}, \ and\ \bibinfo {author} {\bibfnamefont {V.}~\bibnamefont {Vaskonen}},\ }\href@noop {} {\  (\bibinfo {year} {2023})},\ \Eprint {http://arxiv.org/abs/2312.02983} {arXiv:2312.02983 [astro-ph.CO]} \BibitemShut {NoStop}%
\bibitem [{\citenamefont {Ellis}\ \emph {et~al.}(2024{\natexlab{b}})\citenamefont {Ellis}, \citenamefont {Fairbairn}, \citenamefont {H\"utsi}, \citenamefont {Urrutia}, \citenamefont {Vaskonen},\ and\ \citenamefont {Veerm\"ae}}]{Ellis:2024wdh}%
  \BibitemOpen
  \bibfield  {author} {\bibinfo {author} {\bibfnamefont {J.}~\bibnamefont {Ellis}}, \bibinfo {author} {\bibfnamefont {M.}~\bibnamefont {Fairbairn}}, \bibinfo {author} {\bibfnamefont {G.}~\bibnamefont {H\"utsi}}, \bibinfo {author} {\bibfnamefont {J.}~\bibnamefont {Urrutia}}, \bibinfo {author} {\bibfnamefont {V.}~\bibnamefont {Vaskonen}}, \ and\ \bibinfo {author} {\bibfnamefont {H.}~\bibnamefont {Veerm\"ae}},\ }\href@noop {} {\  (\bibinfo {year} {2024}{\natexlab{b}})},\ \Eprint {http://arxiv.org/abs/2403.19650} {arXiv:2403.19650 [astro-ph.CO]} \BibitemShut {NoStop}%
\bibitem [{\citenamefont {Volonteri}\ \emph {et~al.}(2021)\citenamefont {Volonteri}, \citenamefont {Habouzit},\ and\ \citenamefont {Colpi}}]{Volonteri_2021}%
  \BibitemOpen
  \bibfield  {author} {\bibinfo {author} {\bibfnamefont {M.}~\bibnamefont {Volonteri}}, \bibinfo {author} {\bibfnamefont {M.}~\bibnamefont {Habouzit}}, \ and\ \bibinfo {author} {\bibfnamefont {M.}~\bibnamefont {Colpi}},\ }\href {\doibase 10.1038/s42254-021-00364-9} {\bibfield  {journal} {\bibinfo  {journal} {Nature Reviews Physics}\ }\textbf {\bibinfo {volume} {3}},\ \bibinfo {pages} {732–743} (\bibinfo {year} {2021})}\BibitemShut {NoStop}%
\bibitem [{\citenamefont {Larson}\ \emph {et~al.}(2023)\citenamefont {Larson} \emph {et~al.}}]{CEERSTeam:2023qgy}%
  \BibitemOpen
  \bibfield  {author} {\bibinfo {author} {\bibfnamefont {R.~L.}\ \bibnamefont {Larson}} \emph {et~al.} (\bibinfo {collaboration} {CEERS Team}),\ }\href {\doibase 10.3847/2041-8213/ace619} {\bibfield  {journal} {\bibinfo  {journal} {Astrophys. J. Lett.}\ }\textbf {\bibinfo {volume} {953}},\ \bibinfo {pages} {L29} (\bibinfo {year} {2023})},\ \Eprint {http://arxiv.org/abs/2303.08918} {arXiv:2303.08918 [astro-ph.GA]} \BibitemShut {NoStop}%
\bibitem [{\citenamefont {Maiolino}\ \emph {et~al.}(2023)\citenamefont {Maiolino} \emph {et~al.}}]{Maiolino:2023zdu}%
  \BibitemOpen
  \bibfield  {author} {\bibinfo {author} {\bibfnamefont {R.}~\bibnamefont {Maiolino}} \emph {et~al.},\ }\href@noop {} {\  (\bibinfo {year} {2023})},\ \Eprint {http://arxiv.org/abs/2305.12492} {arXiv:2305.12492 [astro-ph.GA]} \BibitemShut {NoStop}%
\bibitem [{\citenamefont {Bogdan}\ \emph {et~al.}(2024)\citenamefont {Bogdan} \emph {et~al.}}]{Bogdan:2023ilu}%
  \BibitemOpen
  \bibfield  {author} {\bibinfo {author} {\bibfnamefont {A.}~\bibnamefont {Bogdan}} \emph {et~al.},\ }\href {\doibase 10.1038/s41550-023-02111-9} {\bibfield  {journal} {\bibinfo  {journal} {Nature Astron.}\ }\textbf {\bibinfo {volume} {8}},\ \bibinfo {pages} {126} (\bibinfo {year} {2024})},\ \Eprint {http://arxiv.org/abs/2305.15458} {arXiv:2305.15458 [astro-ph.GA]} \BibitemShut {NoStop}%
\bibitem [{\citenamefont {Natarajan}\ \emph {et~al.}(2024)\citenamefont {Natarajan}, \citenamefont {Pacucci}, \citenamefont {Ricarte}, \citenamefont {Bogdan}, \citenamefont {Goulding},\ and\ \citenamefont {Cappelluti}}]{Natarajan:2023rxq}%
  \BibitemOpen
  \bibfield  {author} {\bibinfo {author} {\bibfnamefont {P.}~\bibnamefont {Natarajan}}, \bibinfo {author} {\bibfnamefont {F.}~\bibnamefont {Pacucci}}, \bibinfo {author} {\bibfnamefont {A.}~\bibnamefont {Ricarte}}, \bibinfo {author} {\bibfnamefont {A.}~\bibnamefont {Bogdan}}, \bibinfo {author} {\bibfnamefont {A.~D.}\ \bibnamefont {Goulding}}, \ and\ \bibinfo {author} {\bibfnamefont {N.}~\bibnamefont {Cappelluti}},\ }\href {\doibase 10.3847/2041-8213/ad0e76} {\bibfield  {journal} {\bibinfo  {journal} {Astrophys. J. Lett.}\ }\textbf {\bibinfo {volume} {960}},\ \bibinfo {pages} {L1} (\bibinfo {year} {2024})},\ \Eprint {http://arxiv.org/abs/2308.02654} {arXiv:2308.02654 [astro-ph.HE]} \BibitemShut {NoStop}%
\bibitem [{\citenamefont {{Harikane}}\ \emph {et~al.}(2023)\citenamefont {{Harikane}}, \citenamefont {{Zhang}}, \citenamefont {{Nakajima}}, \citenamefont {{Ouchi}}, \citenamefont {{Isobe}}, \citenamefont {{Ono}}, \citenamefont {{Hatano}}, \citenamefont {{Xu}},\ and\ \citenamefont {{Umeda}}}]{2023ApJ...959...39H}%
  \BibitemOpen
  \bibfield  {author} {\bibinfo {author} {\bibfnamefont {Y.}~\bibnamefont {{Harikane}}}, \bibinfo {author} {\bibfnamefont {Y.}~\bibnamefont {{Zhang}}}, \bibinfo {author} {\bibfnamefont {K.}~\bibnamefont {{Nakajima}}}, \bibinfo {author} {\bibfnamefont {M.}~\bibnamefont {{Ouchi}}}, \bibinfo {author} {\bibfnamefont {Y.}~\bibnamefont {{Isobe}}}, \bibinfo {author} {\bibfnamefont {Y.}~\bibnamefont {{Ono}}}, \bibinfo {author} {\bibfnamefont {S.}~\bibnamefont {{Hatano}}}, \bibinfo {author} {\bibfnamefont {Y.}~\bibnamefont {{Xu}}}, \ and\ \bibinfo {author} {\bibfnamefont {H.}~\bibnamefont {{Umeda}}},\ }\href {\doibase 10.3847/1538-4357/ad029e} {\bibfield  {journal} {\bibinfo  {journal} {\apj}\ }\textbf {\bibinfo {volume} {959}},\ \bibinfo {eid} {39} (\bibinfo {year} {2023})},\ \Eprint {http://arxiv.org/abs/2303.11946} {arXiv:2303.11946 [astro-ph.GA]} \BibitemShut {NoStop}%
\bibitem [{\citenamefont {Massonneau}\ \emph {et~al.}(2023)\citenamefont {Massonneau}, \citenamefont {Volonteri}, \citenamefont {Dubois},\ and\ \citenamefont {Beckmann}}]{Massonneau:2022sye}%
  \BibitemOpen
  \bibfield  {author} {\bibinfo {author} {\bibfnamefont {W.}~\bibnamefont {Massonneau}}, \bibinfo {author} {\bibfnamefont {M.}~\bibnamefont {Volonteri}}, \bibinfo {author} {\bibfnamefont {Y.}~\bibnamefont {Dubois}}, \ and\ \bibinfo {author} {\bibfnamefont {R.~S.}\ \bibnamefont {Beckmann}},\ }\href {\doibase 10.1051/0004-6361/202243170} {\bibfield  {journal} {\bibinfo  {journal} {Astron. Astrophys.}\ }\textbf {\bibinfo {volume} {670}},\ \bibinfo {pages} {A180} (\bibinfo {year} {2023})},\ \Eprint {http://arxiv.org/abs/2201.08766} {arXiv:2201.08766 [astro-ph.GA]} \BibitemShut {NoStop}%
\bibitem [{\citenamefont {Banik}\ \emph {et~al.}(2019)\citenamefont {Banik}, \citenamefont {Tan},\ and\ \citenamefont {Monaco}}]{Banik:2016qww}%
  \BibitemOpen
  \bibfield  {author} {\bibinfo {author} {\bibfnamefont {N.}~\bibnamefont {Banik}}, \bibinfo {author} {\bibfnamefont {J.~C.}\ \bibnamefont {Tan}}, \ and\ \bibinfo {author} {\bibfnamefont {P.}~\bibnamefont {Monaco}},\ }\href {\doibase 10.1093/mnras/sty3298} {\bibfield  {journal} {\bibinfo  {journal} {Mon. Not. Roy. Astron. Soc.}\ }\textbf {\bibinfo {volume} {483}},\ \bibinfo {pages} {3592} (\bibinfo {year} {2019})},\ \Eprint {http://arxiv.org/abs/1608.04421} {arXiv:1608.04421 [astro-ph.GA]} \BibitemShut {NoStop}%
\bibitem [{\citenamefont {Balaji}\ \emph {et~al.}(2022{\natexlab{a}})\citenamefont {Balaji}, \citenamefont {Silk},\ and\ \citenamefont {Wu}}]{Balaji:2022rsy}%
  \BibitemOpen
  \bibfield  {author} {\bibinfo {author} {\bibfnamefont {S.}~\bibnamefont {Balaji}}, \bibinfo {author} {\bibfnamefont {J.}~\bibnamefont {Silk}}, \ and\ \bibinfo {author} {\bibfnamefont {Y.-P.}\ \bibnamefont {Wu}},\ }\href {\doibase 10.1088/1475-7516/2022/06/008} {\bibfield  {journal} {\bibinfo  {journal} {JCAP}\ }\textbf {\bibinfo {volume} {06}},\ \bibinfo {pages} {008} (\bibinfo {year} {2022}{\natexlab{a}})},\ \Eprint {http://arxiv.org/abs/2202.00700} {arXiv:2202.00700 [astro-ph.CO]} \BibitemShut {NoStop}%
\bibitem [{\citenamefont {Balaji}\ \emph {et~al.}(2022{\natexlab{b}})\citenamefont {Balaji}, \citenamefont {Domenech},\ and\ \citenamefont {Silk}}]{Balaji:2022dbi}%
  \BibitemOpen
  \bibfield  {author} {\bibinfo {author} {\bibfnamefont {S.}~\bibnamefont {Balaji}}, \bibinfo {author} {\bibfnamefont {G.}~\bibnamefont {Domenech}}, \ and\ \bibinfo {author} {\bibfnamefont {J.}~\bibnamefont {Silk}},\ }\href {\doibase 10.1088/1475-7516/2022/09/016} {\bibfield  {journal} {\bibinfo  {journal} {JCAP}\ }\textbf {\bibinfo {volume} {09}},\ \bibinfo {pages} {016} (\bibinfo {year} {2022}{\natexlab{b}})},\ \Eprint {http://arxiv.org/abs/2205.01696} {arXiv:2205.01696 [astro-ph.CO]} \BibitemShut {NoStop}%
\bibitem [{\citenamefont {Qin}\ \emph {et~al.}(2023)\citenamefont {Qin}, \citenamefont {Geller}, \citenamefont {Balaji}, \citenamefont {McDonough},\ and\ \citenamefont {Kaiser}}]{Qin:2023lgo}%
  \BibitemOpen
  \bibfield  {author} {\bibinfo {author} {\bibfnamefont {W.}~\bibnamefont {Qin}}, \bibinfo {author} {\bibfnamefont {S.~R.}\ \bibnamefont {Geller}}, \bibinfo {author} {\bibfnamefont {S.}~\bibnamefont {Balaji}}, \bibinfo {author} {\bibfnamefont {E.}~\bibnamefont {McDonough}}, \ and\ \bibinfo {author} {\bibfnamefont {D.~I.}\ \bibnamefont {Kaiser}},\ }\href {\doibase 10.1103/PhysRevD.108.043508} {\bibfield  {journal} {\bibinfo  {journal} {Phys. Rev. D}\ }\textbf {\bibinfo {volume} {108}},\ \bibinfo {pages} {043508} (\bibinfo {year} {2023})},\ \Eprint {http://arxiv.org/abs/2303.02168} {arXiv:2303.02168 [astro-ph.CO]} \BibitemShut {NoStop}%
\bibitem [{\citenamefont {Geller}\ \emph {et~al.}(2022)\citenamefont {Geller}, \citenamefont {Qin}, \citenamefont {McDonough},\ and\ \citenamefont {Kaiser}}]{Geller:2022nkr}%
  \BibitemOpen
  \bibfield  {author} {\bibinfo {author} {\bibfnamefont {S.~R.}\ \bibnamefont {Geller}}, \bibinfo {author} {\bibfnamefont {W.}~\bibnamefont {Qin}}, \bibinfo {author} {\bibfnamefont {E.}~\bibnamefont {McDonough}}, \ and\ \bibinfo {author} {\bibfnamefont {D.~I.}\ \bibnamefont {Kaiser}},\ }\href {\doibase 10.1103/PhysRevD.106.063535} {\bibfield  {journal} {\bibinfo  {journal} {Phys. Rev. D}\ }\textbf {\bibinfo {volume} {106}},\ \bibinfo {pages} {063535} (\bibinfo {year} {2022})},\ \Eprint {http://arxiv.org/abs/2205.04471} {arXiv:2205.04471 [hep-th]} \BibitemShut {NoStop}%
\bibitem [{\citenamefont {Aghanim}\ \emph {et~al.}(2020)\citenamefont {Aghanim} \emph {et~al.}}]{Planck:2018vyg}%
  \BibitemOpen
  \bibfield  {author} {\bibinfo {author} {\bibfnamefont {N.}~\bibnamefont {Aghanim}} \emph {et~al.} (\bibinfo {collaboration} {Planck}),\ }\href {\doibase 10.1051/0004-6361/201833910} {\bibfield  {journal} {\bibinfo  {journal} {Astron. Astrophys.}\ }\textbf {\bibinfo {volume} {641}},\ \bibinfo {pages} {A6} (\bibinfo {year} {2020})},\ \bibinfo {note} {[Erratum: Astron.Astrophys. 652, C4 (2021)]},\ \Eprint {http://arxiv.org/abs/1807.06209} {arXiv:1807.06209 [astro-ph.CO]} \BibitemShut {NoStop}%
\bibitem [{\citenamefont {Chluba}\ \emph {et~al.}(2012)\citenamefont {Chluba}, \citenamefont {Erickcek},\ and\ \citenamefont {Ben-Dayan}}]{Chluba:2012we}%
  \BibitemOpen
  \bibfield  {author} {\bibinfo {author} {\bibfnamefont {J.}~\bibnamefont {Chluba}}, \bibinfo {author} {\bibfnamefont {A.~L.}\ \bibnamefont {Erickcek}}, \ and\ \bibinfo {author} {\bibfnamefont {I.}~\bibnamefont {Ben-Dayan}},\ }\href {\doibase 10.1088/0004-637X/758/2/76} {\bibfield  {journal} {\bibinfo  {journal} {Astrophys. J.}\ }\textbf {\bibinfo {volume} {758}},\ \bibinfo {pages} {76} (\bibinfo {year} {2012})},\ \Eprint {http://arxiv.org/abs/1203.2681} {arXiv:1203.2681 [astro-ph.CO]} \BibitemShut {NoStop}%
\bibitem [{\citenamefont {Chluba}\ \emph {et~al.}(2015)\citenamefont {Chluba}, \citenamefont {Hamann},\ and\ \citenamefont {Patil}}]{Chluba:2015bqa}%
  \BibitemOpen
  \bibfield  {author} {\bibinfo {author} {\bibfnamefont {J.}~\bibnamefont {Chluba}}, \bibinfo {author} {\bibfnamefont {J.}~\bibnamefont {Hamann}}, \ and\ \bibinfo {author} {\bibfnamefont {S.~P.}\ \bibnamefont {Patil}},\ }\href {\doibase 10.1142/S0218271815300232} {\bibfield  {journal} {\bibinfo  {journal} {Int. J. Mod. Phys. D}\ }\textbf {\bibinfo {volume} {24}},\ \bibinfo {pages} {1530023} (\bibinfo {year} {2015})},\ \Eprint {http://arxiv.org/abs/1505.01834} {arXiv:1505.01834 [astro-ph.CO]} \BibitemShut {NoStop}%
\bibitem [{\citenamefont {Josan}\ \emph {et~al.}(2009)\citenamefont {Josan}, \citenamefont {Green},\ and\ \citenamefont {Malik}}]{Josan:2009qn}%
  \BibitemOpen
  \bibfield  {author} {\bibinfo {author} {\bibfnamefont {A.~S.}\ \bibnamefont {Josan}}, \bibinfo {author} {\bibfnamefont {A.~M.}\ \bibnamefont {Green}}, \ and\ \bibinfo {author} {\bibfnamefont {K.~A.}\ \bibnamefont {Malik}},\ }\href {\doibase 10.1103/PhysRevD.79.103520} {\bibfield  {journal} {\bibinfo  {journal} {Phys. Rev. D}\ }\textbf {\bibinfo {volume} {79}},\ \bibinfo {pages} {103520} (\bibinfo {year} {2009})},\ \Eprint {http://arxiv.org/abs/0903.3184} {arXiv:0903.3184 [astro-ph.CO]} \BibitemShut {NoStop}%
\bibitem [{\citenamefont {Carr}\ \emph {et~al.}(2010)\citenamefont {Carr}, \citenamefont {Kohri}, \citenamefont {Sendouda},\ and\ \citenamefont {Yokoyama}}]{Carr:2009jm}%
  \BibitemOpen
  \bibfield  {author} {\bibinfo {author} {\bibfnamefont {B.~J.}\ \bibnamefont {Carr}}, \bibinfo {author} {\bibfnamefont {K.}~\bibnamefont {Kohri}}, \bibinfo {author} {\bibfnamefont {Y.}~\bibnamefont {Sendouda}}, \ and\ \bibinfo {author} {\bibfnamefont {J.}~\bibnamefont {Yokoyama}},\ }\href {\doibase 10.1103/PhysRevD.81.104019} {\bibfield  {journal} {\bibinfo  {journal} {Phys. Rev. D}\ }\textbf {\bibinfo {volume} {81}},\ \bibinfo {pages} {104019} (\bibinfo {year} {2010})},\ \Eprint {http://arxiv.org/abs/0912.5297} {arXiv:0912.5297 [astro-ph.CO]} \BibitemShut {NoStop}%
\bibitem [{\citenamefont {Byrnes}\ \emph {et~al.}(2019)\citenamefont {Byrnes}, \citenamefont {Cole},\ and\ \citenamefont {Patil}}]{Byrnes:2018txb}%
  \BibitemOpen
  \bibfield  {author} {\bibinfo {author} {\bibfnamefont {C.~T.}\ \bibnamefont {Byrnes}}, \bibinfo {author} {\bibfnamefont {P.~S.}\ \bibnamefont {Cole}}, \ and\ \bibinfo {author} {\bibfnamefont {S.~P.}\ \bibnamefont {Patil}},\ }\href {\doibase 10.1088/1475-7516/2019/06/028} {\bibfield  {journal} {\bibinfo  {journal} {JCAP}\ }\textbf {\bibinfo {volume} {06}},\ \bibinfo {pages} {028} (\bibinfo {year} {2019})},\ \Eprint {http://arxiv.org/abs/1811.11158} {arXiv:1811.11158 [astro-ph.CO]} \BibitemShut {NoStop}%
\bibitem [{\citenamefont {Dalianis}(2019)}]{Dalianis:2018ymb}%
  \BibitemOpen
  \bibfield  {author} {\bibinfo {author} {\bibfnamefont {I.}~\bibnamefont {Dalianis}},\ }\href {\doibase 10.1088/1475-7516/2019/08/032} {\bibfield  {journal} {\bibinfo  {journal} {JCAP}\ }\textbf {\bibinfo {volume} {08}},\ \bibinfo {pages} {032} (\bibinfo {year} {2019})},\ \Eprint {http://arxiv.org/abs/1812.09807} {arXiv:1812.09807 [astro-ph.CO]} \BibitemShut {NoStop}%
\bibitem [{\citenamefont {Sato-Polito}\ \emph {et~al.}(2019)\citenamefont {Sato-Polito}, \citenamefont {Kovetz},\ and\ \citenamefont {Kamionkowski}}]{Sato-Polito:2019hws}%
  \BibitemOpen
  \bibfield  {author} {\bibinfo {author} {\bibfnamefont {G.}~\bibnamefont {Sato-Polito}}, \bibinfo {author} {\bibfnamefont {E.~D.}\ \bibnamefont {Kovetz}}, \ and\ \bibinfo {author} {\bibfnamefont {M.}~\bibnamefont {Kamionkowski}},\ }\href {\doibase 10.1103/PhysRevD.100.063521} {\bibfield  {journal} {\bibinfo  {journal} {Phys. Rev. D}\ }\textbf {\bibinfo {volume} {100}},\ \bibinfo {pages} {063521} (\bibinfo {year} {2019})},\ \Eprint {http://arxiv.org/abs/1904.10971} {arXiv:1904.10971 [astro-ph.CO]} \BibitemShut {NoStop}%
\bibitem [{\citenamefont {Gow}\ \emph {et~al.}(2021)\citenamefont {Gow}, \citenamefont {Byrnes}, \citenamefont {Cole},\ and\ \citenamefont {Young}}]{Gow:2020bzo}%
  \BibitemOpen
  \bibfield  {author} {\bibinfo {author} {\bibfnamefont {A.~D.}\ \bibnamefont {Gow}}, \bibinfo {author} {\bibfnamefont {C.~T.}\ \bibnamefont {Byrnes}}, \bibinfo {author} {\bibfnamefont {P.~S.}\ \bibnamefont {Cole}}, \ and\ \bibinfo {author} {\bibfnamefont {S.}~\bibnamefont {Young}},\ }\href {\doibase 10.1088/1475-7516/2021/02/002} {\bibfield  {journal} {\bibinfo  {journal} {JCAP}\ }\textbf {\bibinfo {volume} {02}},\ \bibinfo {pages} {002} (\bibinfo {year} {2021})},\ \Eprint {http://arxiv.org/abs/2008.03289} {arXiv:2008.03289 [astro-ph.CO]} \BibitemShut {NoStop}%
\bibitem [{\citenamefont {Delos}\ \emph {et~al.}(2018)\citenamefont {Delos}, \citenamefont {Erickcek}, \citenamefont {Bailey},\ and\ \citenamefont {Alvarez}}]{Delos:2018ueo}%
  \BibitemOpen
  \bibfield  {author} {\bibinfo {author} {\bibfnamefont {M.~S.}\ \bibnamefont {Delos}}, \bibinfo {author} {\bibfnamefont {A.~L.}\ \bibnamefont {Erickcek}}, \bibinfo {author} {\bibfnamefont {A.~P.}\ \bibnamefont {Bailey}}, \ and\ \bibinfo {author} {\bibfnamefont {M.~A.}\ \bibnamefont {Alvarez}},\ }\href {\doibase 10.1103/PhysRevD.98.063527} {\bibfield  {journal} {\bibinfo  {journal} {Phys. Rev. D}\ }\textbf {\bibinfo {volume} {98}},\ \bibinfo {pages} {063527} (\bibinfo {year} {2018})},\ \Eprint {http://arxiv.org/abs/1806.07389} {arXiv:1806.07389 [astro-ph.CO]} \BibitemShut {NoStop}%
\bibitem [{\citenamefont {Nakama}\ \emph {et~al.}(2018)\citenamefont {Nakama}, \citenamefont {Suyama}, \citenamefont {Kohri},\ and\ \citenamefont {Hiroshima}}]{Nakama:2017qac}%
  \BibitemOpen
  \bibfield  {author} {\bibinfo {author} {\bibfnamefont {T.}~\bibnamefont {Nakama}}, \bibinfo {author} {\bibfnamefont {T.}~\bibnamefont {Suyama}}, \bibinfo {author} {\bibfnamefont {K.}~\bibnamefont {Kohri}}, \ and\ \bibinfo {author} {\bibfnamefont {N.}~\bibnamefont {Hiroshima}},\ }\href {\doibase 10.1103/PhysRevD.97.023539} {\bibfield  {journal} {\bibinfo  {journal} {Phys. Rev. D}\ }\textbf {\bibinfo {volume} {97}},\ \bibinfo {pages} {023539} (\bibinfo {year} {2018})},\ \Eprint {http://arxiv.org/abs/1712.08820} {arXiv:1712.08820 [astro-ph.CO]} \BibitemShut {NoStop}%
\bibitem [{\citenamefont {Abe}\ \emph {et~al.}(2022)\citenamefont {Abe}, \citenamefont {Minoda},\ and\ \citenamefont {Tashiro}}]{Abe:2021mcv}%
  \BibitemOpen
  \bibfield  {author} {\bibinfo {author} {\bibfnamefont {K.~T.}\ \bibnamefont {Abe}}, \bibinfo {author} {\bibfnamefont {T.}~\bibnamefont {Minoda}}, \ and\ \bibinfo {author} {\bibfnamefont {H.}~\bibnamefont {Tashiro}},\ }\href {\doibase 10.1103/PhysRevD.105.063531} {\bibfield  {journal} {\bibinfo  {journal} {Phys. Rev. D}\ }\textbf {\bibinfo {volume} {105}},\ \bibinfo {pages} {063531} (\bibinfo {year} {2022})},\ \Eprint {http://arxiv.org/abs/2108.00621} {arXiv:2108.00621 [astro-ph.CO]} \BibitemShut {NoStop}%
\bibitem [{\citenamefont {Yoshiura}\ \emph {et~al.}(2020)\citenamefont {Yoshiura}, \citenamefont {Oguri}, \citenamefont {Takahashi},\ and\ \citenamefont {Takahashi}}]{Yoshiura:2020soa}%
  \BibitemOpen
  \bibfield  {author} {\bibinfo {author} {\bibfnamefont {S.}~\bibnamefont {Yoshiura}}, \bibinfo {author} {\bibfnamefont {M.}~\bibnamefont {Oguri}}, \bibinfo {author} {\bibfnamefont {K.}~\bibnamefont {Takahashi}}, \ and\ \bibinfo {author} {\bibfnamefont {T.}~\bibnamefont {Takahashi}},\ }\href {\doibase 10.1103/PhysRevD.102.083515} {\bibfield  {journal} {\bibinfo  {journal} {Phys. Rev. D}\ }\textbf {\bibinfo {volume} {102}},\ \bibinfo {pages} {083515} (\bibinfo {year} {2020})},\ \Eprint {http://arxiv.org/abs/2007.14695} {arXiv:2007.14695 [astro-ph.CO]} \BibitemShut {NoStop}%
\bibitem [{\citenamefont {Gilman}\ \emph {et~al.}(2022)\citenamefont {Gilman}, \citenamefont {Benson}, \citenamefont {Bovy}, \citenamefont {Birrer}, \citenamefont {Treu},\ and\ \citenamefont {Nierenberg}}]{Gilman:2021gkj}%
  \BibitemOpen
  \bibfield  {author} {\bibinfo {author} {\bibfnamefont {D.}~\bibnamefont {Gilman}}, \bibinfo {author} {\bibfnamefont {A.}~\bibnamefont {Benson}}, \bibinfo {author} {\bibfnamefont {J.}~\bibnamefont {Bovy}}, \bibinfo {author} {\bibfnamefont {S.}~\bibnamefont {Birrer}}, \bibinfo {author} {\bibfnamefont {T.}~\bibnamefont {Treu}}, \ and\ \bibinfo {author} {\bibfnamefont {A.}~\bibnamefont {Nierenberg}},\ }\href {\doibase 10.1093/mnras/stac670} {\bibfield  {journal} {\bibinfo  {journal} {Mon. Not. Roy. Astron. Soc.}\ }\textbf {\bibinfo {volume} {512}},\ \bibinfo {pages} {3163} (\bibinfo {year} {2022})},\ \Eprint {http://arxiv.org/abs/2112.03293} {arXiv:2112.03293 [astro-ph.CO]} \BibitemShut {NoStop}%
\bibitem [{\citenamefont {\"Unal}\ \emph {et~al.}(2021)\citenamefont {\"Unal}, \citenamefont {Kovetz},\ and\ \citenamefont {Patil}}]{Unal:2020mts}%
  \BibitemOpen
  \bibfield  {author} {\bibinfo {author} {\bibfnamefont {C.}~\bibnamefont {\"Unal}}, \bibinfo {author} {\bibfnamefont {E.~D.}\ \bibnamefont {Kovetz}}, \ and\ \bibinfo {author} {\bibfnamefont {S.~P.}\ \bibnamefont {Patil}},\ }\href {\doibase 10.1103/PhysRevD.103.063519} {\bibfield  {journal} {\bibinfo  {journal} {Phys. Rev. D}\ }\textbf {\bibinfo {volume} {103}},\ \bibinfo {pages} {063519} (\bibinfo {year} {2021})},\ \Eprint {http://arxiv.org/abs/2008.11184} {arXiv:2008.11184 [astro-ph.CO]} \BibitemShut {NoStop}%
\bibitem [{\citenamefont {Iovino}\ \emph {et~al.}(2024)\citenamefont {Iovino}, \citenamefont {Perna}, \citenamefont {Riotto},\ and\ \citenamefont {Veerm\"ae}}]{Iovino:2024uxp}%
  \BibitemOpen
  \bibfield  {author} {\bibinfo {author} {\bibfnamefont {A.~J.}\ \bibnamefont {Iovino}}, \bibinfo {author} {\bibfnamefont {G.}~\bibnamefont {Perna}}, \bibinfo {author} {\bibfnamefont {A.}~\bibnamefont {Riotto}}, \ and\ \bibinfo {author} {\bibfnamefont {H.}~\bibnamefont {Veerm\"ae}},\ }\href@noop {} {\  (\bibinfo {year} {2024})},\ \Eprint {http://arxiv.org/abs/2406.20089} {arXiv:2406.20089 [astro-ph.CO]} \BibitemShut {NoStop}%
\bibitem [{\citenamefont {Drlica-Wagner}\ \emph {et~al.}(2020)\citenamefont {Drlica-Wagner} \emph {et~al.}}]{DES:2019vzn}%
  \BibitemOpen
  \bibfield  {author} {\bibinfo {author} {\bibfnamefont {A.}~\bibnamefont {Drlica-Wagner}} \emph {et~al.} (\bibinfo {collaboration} {DES}),\ }\href {\doibase 10.3847/1538-4357/ab7eb9} {\bibfield  {journal} {\bibinfo  {journal} {Astrophys. J.}\ }\textbf {\bibinfo {volume} {893}},\ \bibinfo {pages} {47} (\bibinfo {year} {2020})},\ \Eprint {http://arxiv.org/abs/1912.03302} {arXiv:1912.03302 [astro-ph.GA]} \BibitemShut {NoStop}%
\bibitem [{\citenamefont {Grillmair}\ and\ \citenamefont {Dionatos}(2006)}]{Grillmair:2006bd}%
  \BibitemOpen
  \bibfield  {author} {\bibinfo {author} {\bibfnamefont {C.~J.}\ \bibnamefont {Grillmair}}\ and\ \bibinfo {author} {\bibfnamefont {O.}~\bibnamefont {Dionatos}},\ }\href {\doibase 10.1086/505111} {\bibfield  {journal} {\bibinfo  {journal} {Astrophys. J. Lett.}\ }\textbf {\bibinfo {volume} {643}},\ \bibinfo {pages} {L17} (\bibinfo {year} {2006})},\ \Eprint {http://arxiv.org/abs/astro-ph/0604332} {arXiv:astro-ph/0604332} \BibitemShut {NoStop}%
\bibitem [{\citenamefont {Ibata}\ \emph {et~al.}(2002)\citenamefont {Ibata}, \citenamefont {Lewis},\ and\ \citenamefont {Irwin}}]{Ibata:2001iv}%
  \BibitemOpen
  \bibfield  {author} {\bibinfo {author} {\bibfnamefont {R.~A.}\ \bibnamefont {Ibata}}, \bibinfo {author} {\bibfnamefont {G.~F.}\ \bibnamefont {Lewis}}, \ and\ \bibinfo {author} {\bibfnamefont {M.~J.}\ \bibnamefont {Irwin}},\ }\href {\doibase 10.1046/j.1365-8711.2002.05358.x} {\bibfield  {journal} {\bibinfo  {journal} {Mon. Not. Roy. Astron. Soc.}\ }\textbf {\bibinfo {volume} {332}},\ \bibinfo {pages} {915} (\bibinfo {year} {2002})},\ \Eprint {http://arxiv.org/abs/astro-ph/0110690} {arXiv:astro-ph/0110690} \BibitemShut {NoStop}%
\bibitem [{\citenamefont {Johnston}\ \emph {et~al.}(2002)\citenamefont {Johnston}, \citenamefont {Spergel},\ and\ \citenamefont {Haydn}}]{Johnston:2001wh}%
  \BibitemOpen
  \bibfield  {author} {\bibinfo {author} {\bibfnamefont {K.~V.}\ \bibnamefont {Johnston}}, \bibinfo {author} {\bibfnamefont {D.~N.}\ \bibnamefont {Spergel}}, \ and\ \bibinfo {author} {\bibfnamefont {C.}~\bibnamefont {Haydn}},\ }\href {\doibase 10.1086/339791} {\bibfield  {journal} {\bibinfo  {journal} {Astrophys. J.}\ }\textbf {\bibinfo {volume} {570}},\ \bibinfo {pages} {656} (\bibinfo {year} {2002})},\ \Eprint {http://arxiv.org/abs/astro-ph/0111196} {arXiv:astro-ph/0111196} \BibitemShut {NoStop}%
\bibitem [{\citenamefont {Siegal-Gaskins}\ and\ \citenamefont {Valluri}(2008)}]{Siegal-Gaskins:2007zos}%
  \BibitemOpen
  \bibfield  {author} {\bibinfo {author} {\bibfnamefont {J.~M.}\ \bibnamefont {Siegal-Gaskins}}\ and\ \bibinfo {author} {\bibfnamefont {M.}~\bibnamefont {Valluri}},\ }\href {\doibase 10.1086/587450} {\bibfield  {journal} {\bibinfo  {journal} {Astrophys. J.}\ }\textbf {\bibinfo {volume} {681}},\ \bibinfo {pages} {40} (\bibinfo {year} {2008})},\ \Eprint {http://arxiv.org/abs/0710.0385} {arXiv:0710.0385 [astro-ph]} \BibitemShut {NoStop}%
\bibitem [{\citenamefont {Carlberg}(2009)}]{Carlberg:2009ae}%
  \BibitemOpen
  \bibfield  {author} {\bibinfo {author} {\bibfnamefont {R.~G.}\ \bibnamefont {Carlberg}},\ }\href {\doibase 10.1088/0004-637X/705/2/L223} {\bibfield  {journal} {\bibinfo  {journal} {Astrophys. J. Lett.}\ }\textbf {\bibinfo {volume} {705}},\ \bibinfo {pages} {L223} (\bibinfo {year} {2009})},\ \Eprint {http://arxiv.org/abs/0908.4345} {arXiv:0908.4345 [astro-ph.CO]} \BibitemShut {NoStop}%
\bibitem [{\citenamefont {Nadler}\ \emph {et~al.}(2021)\citenamefont {Nadler}, \citenamefont {Birrer}, \citenamefont {Gilman}, \citenamefont {Wechsler}, \citenamefont {Du}, \citenamefont {Benson}, \citenamefont {Nierenberg},\ and\ \citenamefont {Treu}}]{Nadler:2021dft}%
  \BibitemOpen
  \bibfield  {author} {\bibinfo {author} {\bibfnamefont {E.~O.}\ \bibnamefont {Nadler}}, \bibinfo {author} {\bibfnamefont {S.}~\bibnamefont {Birrer}}, \bibinfo {author} {\bibfnamefont {D.}~\bibnamefont {Gilman}}, \bibinfo {author} {\bibfnamefont {R.~H.}\ \bibnamefont {Wechsler}}, \bibinfo {author} {\bibfnamefont {X.}~\bibnamefont {Du}}, \bibinfo {author} {\bibfnamefont {A.}~\bibnamefont {Benson}}, \bibinfo {author} {\bibfnamefont {A.~M.}\ \bibnamefont {Nierenberg}}, \ and\ \bibinfo {author} {\bibfnamefont {T.}~\bibnamefont {Treu}},\ }\href {\doibase 10.3847/1538-4357/abf9a3} {\bibfield  {journal} {\bibinfo  {journal} {Astrophys. J.}\ }\textbf {\bibinfo {volume} {917}},\ \bibinfo {pages} {7} (\bibinfo {year} {2021})},\ \Eprint {http://arxiv.org/abs/2101.07810} {arXiv:2101.07810 [astro-ph.CO]} \BibitemShut {NoStop}%
\bibitem [{\citenamefont {Banik}\ \emph {et~al.}(2021{\natexlab{a}})\citenamefont {Banik}, \citenamefont {Bovy}, \citenamefont {Bertone}, \citenamefont {Erkal},\ and\ \citenamefont {de~Boer}}]{Banik:2019cza}%
  \BibitemOpen
  \bibfield  {author} {\bibinfo {author} {\bibfnamefont {N.}~\bibnamefont {Banik}}, \bibinfo {author} {\bibfnamefont {J.}~\bibnamefont {Bovy}}, \bibinfo {author} {\bibfnamefont {G.}~\bibnamefont {Bertone}}, \bibinfo {author} {\bibfnamefont {D.}~\bibnamefont {Erkal}}, \ and\ \bibinfo {author} {\bibfnamefont {T.~J.~L.}\ \bibnamefont {de~Boer}},\ }\href {\doibase 10.1093/mnras/stab210} {\bibfield  {journal} {\bibinfo  {journal} {Mon. Not. Roy. Astron. Soc.}\ }\textbf {\bibinfo {volume} {502}},\ \bibinfo {pages} {2364} (\bibinfo {year} {2021}{\natexlab{a}})},\ \Eprint {http://arxiv.org/abs/1911.02662} {arXiv:1911.02662 [astro-ph.GA]} \BibitemShut {NoStop}%
\bibitem [{\citenamefont {Banik}\ \emph {et~al.}(2021{\natexlab{b}})\citenamefont {Banik}, \citenamefont {Bovy}, \citenamefont {Bertone}, \citenamefont {Erkal},\ and\ \citenamefont {de~Boer}}]{Banik:2019smi}%
  \BibitemOpen
  \bibfield  {author} {\bibinfo {author} {\bibfnamefont {N.}~\bibnamefont {Banik}}, \bibinfo {author} {\bibfnamefont {J.}~\bibnamefont {Bovy}}, \bibinfo {author} {\bibfnamefont {G.}~\bibnamefont {Bertone}}, \bibinfo {author} {\bibfnamefont {D.}~\bibnamefont {Erkal}}, \ and\ \bibinfo {author} {\bibfnamefont {T.~J.~L.}\ \bibnamefont {de~Boer}},\ }\href {\doibase 10.1088/1475-7516/2021/10/043} {\bibfield  {journal} {\bibinfo  {journal} {JCAP}\ }\textbf {\bibinfo {volume} {10}},\ \bibinfo {pages} {043} (\bibinfo {year} {2021}{\natexlab{b}})},\ \Eprint {http://arxiv.org/abs/1911.02663} {arXiv:1911.02663 [astro-ph.GA]} \BibitemShut {NoStop}%
\bibitem [{\citenamefont {Dekker}\ \emph {et~al.}(2021)\citenamefont {Dekker}, \citenamefont {Ando}, \citenamefont {Correa},\ and\ \citenamefont {Ng}}]{Dekker:2021scf}%
  \BibitemOpen
  \bibfield  {author} {\bibinfo {author} {\bibfnamefont {A.}~\bibnamefont {Dekker}}, \bibinfo {author} {\bibfnamefont {S.}~\bibnamefont {Ando}}, \bibinfo {author} {\bibfnamefont {C.~A.}\ \bibnamefont {Correa}}, \ and\ \bibinfo {author} {\bibfnamefont {K.~C.~Y.}\ \bibnamefont {Ng}},\ }\href@noop {} {\  (\bibinfo {year} {2021})},\ \Eprint {http://arxiv.org/abs/2111.13137} {arXiv:2111.13137 [astro-ph.CO]} \BibitemShut {NoStop}%
\bibitem [{\citenamefont {Ozsoy}\ and\ \citenamefont {Tasinato}(2020)}]{Ozsoy:2019lyy}%
  \BibitemOpen
  \bibfield  {author} {\bibinfo {author} {\bibfnamefont {O.}~\bibnamefont {Ozsoy}}\ and\ \bibinfo {author} {\bibfnamefont {G.}~\bibnamefont {Tasinato}},\ }\href {\doibase 10.1088/1475-7516/2020/04/048} {\bibfield  {journal} {\bibinfo  {journal} {JCAP}\ }\textbf {\bibinfo {volume} {04}},\ \bibinfo {pages} {048} (\bibinfo {year} {2020})},\ \Eprint {http://arxiv.org/abs/1912.01061} {arXiv:1912.01061 [astro-ph.CO]} \BibitemShut {NoStop}%
\bibitem [{\citenamefont {Carr}\ \emph {et~al.}(2021)\citenamefont {Carr}, \citenamefont {Kohri}, \citenamefont {Sendouda},\ and\ \citenamefont {Yokoyama}}]{Carr:2020gox}%
  \BibitemOpen
  \bibfield  {author} {\bibinfo {author} {\bibfnamefont {B.}~\bibnamefont {Carr}}, \bibinfo {author} {\bibfnamefont {K.}~\bibnamefont {Kohri}}, \bibinfo {author} {\bibfnamefont {Y.}~\bibnamefont {Sendouda}}, \ and\ \bibinfo {author} {\bibfnamefont {J.}~\bibnamefont {Yokoyama}},\ }\href {\doibase 10.1088/1361-6633/ac1e31} {\bibfield  {journal} {\bibinfo  {journal} {Rept. Prog. Phys.}\ }\textbf {\bibinfo {volume} {84}},\ \bibinfo {pages} {116902} (\bibinfo {year} {2021})},\ \Eprint {http://arxiv.org/abs/2002.12778} {arXiv:2002.12778 [astro-ph.CO]} \BibitemShut {NoStop}%
\bibitem [{\citenamefont {Escriv\`a}\ \emph {et~al.}(2022)\citenamefont {Escriv\`a}, \citenamefont {Kuhnel},\ and\ \citenamefont {Tada}}]{Escriva:2022duf}%
  \BibitemOpen
  \bibfield  {author} {\bibinfo {author} {\bibfnamefont {A.}~\bibnamefont {Escriv\`a}}, \bibinfo {author} {\bibfnamefont {F.}~\bibnamefont {Kuhnel}}, \ and\ \bibinfo {author} {\bibfnamefont {Y.}~\bibnamefont {Tada}},\ }\href {\doibase 10.1016/B978-0-32-395636-9.00012-8} {\  (\bibinfo {year} {2022}),\ 10.1016/B978-0-32-395636-9.00012-8},\ \Eprint {http://arxiv.org/abs/2211.05767} {arXiv:2211.05767 [astro-ph.CO]} \BibitemShut {NoStop}%
\bibitem [{\citenamefont {Hooper}\ \emph {et~al.}(2023)\citenamefont {Hooper}, \citenamefont {Ireland}, \citenamefont {Krnjaic},\ and\ \citenamefont {Stebbins}}]{Hooper:2023nnl}%
  \BibitemOpen
  \bibfield  {author} {\bibinfo {author} {\bibfnamefont {D.}~\bibnamefont {Hooper}}, \bibinfo {author} {\bibfnamefont {A.}~\bibnamefont {Ireland}}, \bibinfo {author} {\bibfnamefont {G.}~\bibnamefont {Krnjaic}}, \ and\ \bibinfo {author} {\bibfnamefont {A.}~\bibnamefont {Stebbins}},\ }\href@noop {} {\  (\bibinfo {year} {2023})},\ \Eprint {http://arxiv.org/abs/2308.00756} {arXiv:2308.00756 [astro-ph.CO]} \BibitemShut {NoStop}%
\bibitem [{\citenamefont {{Carr}}(1975)}]{1975ApJ...201....1C}%
  \BibitemOpen
  \bibfield  {author} {\bibinfo {author} {\bibfnamefont {B.~J.}\ \bibnamefont {{Carr}}},\ }\href {\doibase 10.1086/153853} {\bibfield  {journal} {\bibinfo  {journal} {\apj}\ }\textbf {\bibinfo {volume} {201}},\ \bibinfo {pages} {1} (\bibinfo {year} {1975})}\BibitemShut {NoStop}%
\bibitem [{\citenamefont {Germani}\ and\ \citenamefont {Musco}(2019)}]{Germani:2018jgr}%
  \BibitemOpen
  \bibfield  {author} {\bibinfo {author} {\bibfnamefont {C.}~\bibnamefont {Germani}}\ and\ \bibinfo {author} {\bibfnamefont {I.}~\bibnamefont {Musco}},\ }\href {\doibase 10.1103/PhysRevLett.122.141302} {\bibfield  {journal} {\bibinfo  {journal} {Phys. Rev. Lett.}\ }\textbf {\bibinfo {volume} {122}},\ \bibinfo {pages} {141302} (\bibinfo {year} {2019})},\ \Eprint {http://arxiv.org/abs/1805.04087} {arXiv:1805.04087 [astro-ph.CO]} \BibitemShut {NoStop}%
\bibitem [{\citenamefont {Nakama}\ \emph {et~al.}(2017{\natexlab{a}})\citenamefont {Nakama}, \citenamefont {Silk},\ and\ \citenamefont {Kamionkowski}}]{Nakama:2016gzw}%
  \BibitemOpen
  \bibfield  {author} {\bibinfo {author} {\bibfnamefont {T.}~\bibnamefont {Nakama}}, \bibinfo {author} {\bibfnamefont {J.}~\bibnamefont {Silk}}, \ and\ \bibinfo {author} {\bibfnamefont {M.}~\bibnamefont {Kamionkowski}},\ }\href {\doibase 10.1103/PhysRevD.95.043511} {\bibfield  {journal} {\bibinfo  {journal} {Phys. Rev. D}\ }\textbf {\bibinfo {volume} {95}},\ \bibinfo {pages} {043511} (\bibinfo {year} {2017}{\natexlab{a}})},\ \Eprint {http://arxiv.org/abs/1612.06264} {arXiv:1612.06264 [astro-ph.CO]} \BibitemShut {NoStop}%
\bibitem [{\citenamefont {Garcia-Bellido}\ \emph {et~al.}(2017)\citenamefont {Garcia-Bellido}, \citenamefont {Peloso},\ and\ \citenamefont {Unal}}]{Garcia-Bellido:2017aan}%
  \BibitemOpen
  \bibfield  {author} {\bibinfo {author} {\bibfnamefont {J.}~\bibnamefont {Garcia-Bellido}}, \bibinfo {author} {\bibfnamefont {M.}~\bibnamefont {Peloso}}, \ and\ \bibinfo {author} {\bibfnamefont {C.}~\bibnamefont {Unal}},\ }\href {\doibase 10.1088/1475-7516/2017/09/013} {\bibfield  {journal} {\bibinfo  {journal} {JCAP}\ }\textbf {\bibinfo {volume} {09}},\ \bibinfo {pages} {013} (\bibinfo {year} {2017})},\ \Eprint {http://arxiv.org/abs/1707.02441} {arXiv:1707.02441 [astro-ph.CO]} \BibitemShut {NoStop}%
\bibitem [{\citenamefont {Sasaki}\ \emph {et~al.}(2018)\citenamefont {Sasaki}, \citenamefont {Suyama}, \citenamefont {Tanaka},\ and\ \citenamefont {Yokoyama}}]{Sasaki:2018dmp}%
  \BibitemOpen
  \bibfield  {author} {\bibinfo {author} {\bibfnamefont {M.}~\bibnamefont {Sasaki}}, \bibinfo {author} {\bibfnamefont {T.}~\bibnamefont {Suyama}}, \bibinfo {author} {\bibfnamefont {T.}~\bibnamefont {Tanaka}}, \ and\ \bibinfo {author} {\bibfnamefont {S.}~\bibnamefont {Yokoyama}},\ }\href {\doibase 10.1088/1361-6382/aaa7b4} {\bibfield  {journal} {\bibinfo  {journal} {Class. Quant. Grav.}\ }\textbf {\bibinfo {volume} {35}},\ \bibinfo {pages} {063001} (\bibinfo {year} {2018})},\ \Eprint {http://arxiv.org/abs/1801.05235} {arXiv:1801.05235 [astro-ph.CO]} \BibitemShut {NoStop}%
\bibitem [{\citenamefont {Garcia-Bellido}\ \emph {et~al.}(2016)\citenamefont {Garcia-Bellido}, \citenamefont {Peloso},\ and\ \citenamefont {Unal}}]{Garcia-Bellido:2016dkw}%
  \BibitemOpen
  \bibfield  {author} {\bibinfo {author} {\bibfnamefont {J.}~\bibnamefont {Garcia-Bellido}}, \bibinfo {author} {\bibfnamefont {M.}~\bibnamefont {Peloso}}, \ and\ \bibinfo {author} {\bibfnamefont {C.}~\bibnamefont {Unal}},\ }\href {\doibase 10.1088/1475-7516/2016/12/031} {\bibfield  {journal} {\bibinfo  {journal} {JCAP}\ }\textbf {\bibinfo {volume} {12}},\ \bibinfo {pages} {031} (\bibinfo {year} {2016})},\ \Eprint {http://arxiv.org/abs/1610.03763} {arXiv:1610.03763 [astro-ph.CO]} \BibitemShut {NoStop}%
\bibitem [{\citenamefont {Lyth}(2012)}]{Lyth:2012yp}%
  \BibitemOpen
  \bibfield  {author} {\bibinfo {author} {\bibfnamefont {D.~H.}\ \bibnamefont {Lyth}},\ }\href {\doibase 10.1088/1475-7516/2012/05/022} {\bibfield  {journal} {\bibinfo  {journal} {JCAP}\ }\textbf {\bibinfo {volume} {05}},\ \bibinfo {pages} {022} (\bibinfo {year} {2012})},\ \Eprint {http://arxiv.org/abs/1201.4312} {arXiv:1201.4312 [astro-ph.CO]} \BibitemShut {NoStop}%
\bibitem [{\citenamefont {Byrnes}\ \emph {et~al.}(2012)\citenamefont {Byrnes}, \citenamefont {Copeland},\ and\ \citenamefont {Green}}]{Byrnes:2012yx}%
  \BibitemOpen
  \bibfield  {author} {\bibinfo {author} {\bibfnamefont {C.~T.}\ \bibnamefont {Byrnes}}, \bibinfo {author} {\bibfnamefont {E.~J.}\ \bibnamefont {Copeland}}, \ and\ \bibinfo {author} {\bibfnamefont {A.~M.}\ \bibnamefont {Green}},\ }\href {\doibase 10.1103/PhysRevD.86.043512} {\bibfield  {journal} {\bibinfo  {journal} {Phys. Rev. D}\ }\textbf {\bibinfo {volume} {86}},\ \bibinfo {pages} {043512} (\bibinfo {year} {2012})},\ \Eprint {http://arxiv.org/abs/1206.4188} {arXiv:1206.4188 [astro-ph.CO]} \BibitemShut {NoStop}%
\bibitem [{\citenamefont {Kawasaki}\ and\ \citenamefont {Nakatsuka}(2019)}]{Kawasaki:2019mbl}%
  \BibitemOpen
  \bibfield  {author} {\bibinfo {author} {\bibfnamefont {M.}~\bibnamefont {Kawasaki}}\ and\ \bibinfo {author} {\bibfnamefont {H.}~\bibnamefont {Nakatsuka}},\ }\href {\doibase 10.1103/PhysRevD.99.123501} {\bibfield  {journal} {\bibinfo  {journal} {Phys. Rev. D}\ }\textbf {\bibinfo {volume} {99}},\ \bibinfo {pages} {123501} (\bibinfo {year} {2019})},\ \Eprint {http://arxiv.org/abs/1903.02994} {arXiv:1903.02994 [astro-ph.CO]} \BibitemShut {NoStop}%
\bibitem [{\citenamefont {Ando}\ \emph {et~al.}(2018)\citenamefont {Ando}, \citenamefont {Inomata},\ and\ \citenamefont {Kawasaki}}]{Ando:2018qdb}%
  \BibitemOpen
  \bibfield  {author} {\bibinfo {author} {\bibfnamefont {K.}~\bibnamefont {Ando}}, \bibinfo {author} {\bibfnamefont {K.}~\bibnamefont {Inomata}}, \ and\ \bibinfo {author} {\bibfnamefont {M.}~\bibnamefont {Kawasaki}},\ }\href {\doibase 10.1103/PhysRevD.97.103528} {\bibfield  {journal} {\bibinfo  {journal} {Phys. Rev. D}\ }\textbf {\bibinfo {volume} {97}},\ \bibinfo {pages} {103528} (\bibinfo {year} {2018})},\ \Eprint {http://arxiv.org/abs/1802.06393} {arXiv:1802.06393 [astro-ph.CO]} \BibitemShut {NoStop}%
\bibitem [{\citenamefont {Young}\ and\ \citenamefont {Byrnes}(2015)}]{Young:2014oea}%
  \BibitemOpen
  \bibfield  {author} {\bibinfo {author} {\bibfnamefont {S.}~\bibnamefont {Young}}\ and\ \bibinfo {author} {\bibfnamefont {C.~T.}\ \bibnamefont {Byrnes}},\ }\href {\doibase 10.1103/PhysRevD.91.083521} {\bibfield  {journal} {\bibinfo  {journal} {Phys. Rev. D}\ }\textbf {\bibinfo {volume} {91}},\ \bibinfo {pages} {083521} (\bibinfo {year} {2015})},\ \Eprint {http://arxiv.org/abs/1411.4620} {arXiv:1411.4620 [astro-ph.CO]} \BibitemShut {NoStop}%
\bibitem [{\citenamefont {Carr}(1975)}]{Carr:1975qj}%
  \BibitemOpen
  \bibfield  {author} {\bibinfo {author} {\bibfnamefont {B.~J.}\ \bibnamefont {Carr}},\ }\href {\doibase 10.1086/153853} {\bibfield  {journal} {\bibinfo  {journal} {Astrophys. J.}\ }\textbf {\bibinfo {volume} {201}},\ \bibinfo {pages} {1} (\bibinfo {year} {1975})}\BibitemShut {NoStop}%
\bibitem [{\citenamefont {Musco}\ and\ \citenamefont {Miller}(2013)}]{Musco:2012au}%
  \BibitemOpen
  \bibfield  {author} {\bibinfo {author} {\bibfnamefont {I.}~\bibnamefont {Musco}}\ and\ \bibinfo {author} {\bibfnamefont {J.~C.}\ \bibnamefont {Miller}},\ }\href {\doibase 10.1088/0264-9381/30/14/145009} {\bibfield  {journal} {\bibinfo  {journal} {Class. Quant. Grav.}\ }\textbf {\bibinfo {volume} {30}},\ \bibinfo {pages} {145009} (\bibinfo {year} {2013})},\ \Eprint {http://arxiv.org/abs/1201.2379} {arXiv:1201.2379 [gr-qc]} \BibitemShut {NoStop}%
\bibitem [{\citenamefont {Conselice}\ \emph {et~al.}(2016)\citenamefont {Conselice}, \citenamefont {Wilkinson}, \citenamefont {Duncan},\ and\ \citenamefont {Mortlock}}]{Conselice_2016}%
  \BibitemOpen
  \bibfield  {author} {\bibinfo {author} {\bibfnamefont {C.~J.}\ \bibnamefont {Conselice}}, \bibinfo {author} {\bibfnamefont {A.}~\bibnamefont {Wilkinson}}, \bibinfo {author} {\bibfnamefont {K.}~\bibnamefont {Duncan}}, \ and\ \bibinfo {author} {\bibfnamefont {A.}~\bibnamefont {Mortlock}},\ }\href {\doibase 10.3847/0004-637x/830/2/83} {\bibfield  {journal} {\bibinfo  {journal} {The Astrophysical Journal}\ }\textbf {\bibinfo {volume} {830}},\ \bibinfo {pages} {83} (\bibinfo {year} {2016})}\BibitemShut {NoStop}%
\bibitem [{\citenamefont {Ando}\ \emph {et~al.}(2022)\citenamefont {Ando}, \citenamefont {Hiroshima},\ and\ \citenamefont {Ishiwata}}]{Ando:2022tpj}%
  \BibitemOpen
  \bibfield  {author} {\bibinfo {author} {\bibfnamefont {S.}~\bibnamefont {Ando}}, \bibinfo {author} {\bibfnamefont {N.}~\bibnamefont {Hiroshima}}, \ and\ \bibinfo {author} {\bibfnamefont {K.}~\bibnamefont {Ishiwata}},\ }\href {\doibase 10.1103/PhysRevD.106.103014} {\bibfield  {journal} {\bibinfo  {journal} {Phys. Rev. D}\ }\textbf {\bibinfo {volume} {106}},\ \bibinfo {pages} {103014} (\bibinfo {year} {2022})},\ \Eprint {http://arxiv.org/abs/2207.05747} {arXiv:2207.05747 [astro-ph.CO]} \BibitemShut {NoStop}%
\bibitem [{\citenamefont {Schneider}\ \emph {et~al.}(2013)\citenamefont {Schneider}, \citenamefont {Smith},\ and\ \citenamefont {Reed}}]{Schneider:2013ria}%
  \BibitemOpen
  \bibfield  {author} {\bibinfo {author} {\bibfnamefont {A.}~\bibnamefont {Schneider}}, \bibinfo {author} {\bibfnamefont {R.~E.}\ \bibnamefont {Smith}}, \ and\ \bibinfo {author} {\bibfnamefont {D.}~\bibnamefont {Reed}},\ }\href {\doibase 10.1093/mnras/stt829} {\bibfield  {journal} {\bibinfo  {journal} {Mon. Not. Roy. Astron. Soc.}\ }\textbf {\bibinfo {volume} {433}},\ \bibinfo {pages} {1573} (\bibinfo {year} {2013})},\ \Eprint {http://arxiv.org/abs/1303.0839} {arXiv:1303.0839 [astro-ph.CO]} \BibitemShut {NoStop}%
\bibitem [{\citenamefont {Yang}\ \emph {et~al.}(2011)\citenamefont {Yang}, \citenamefont {Mo}, \citenamefont {Zhang},\ and\ \citenamefont {Bosch}}]{Yang:2011rf}%
  \BibitemOpen
  \bibfield  {author} {\bibinfo {author} {\bibfnamefont {X.}~\bibnamefont {Yang}}, \bibinfo {author} {\bibfnamefont {H.~J.}\ \bibnamefont {Mo}}, \bibinfo {author} {\bibfnamefont {Y.}~\bibnamefont {Zhang}}, \ and\ \bibinfo {author} {\bibfnamefont {F.~C. v.~d.}\ \bibnamefont {Bosch}},\ }\href {\doibase 10.1088/0004-637X/741/1/13} {\bibfield  {journal} {\bibinfo  {journal} {Astrophys. J.}\ }\textbf {\bibinfo {volume} {741}},\ \bibinfo {pages} {13} (\bibinfo {year} {2011})},\ \Eprint {http://arxiv.org/abs/1104.1757} {arXiv:1104.1757 [astro-ph.CO]} \BibitemShut {NoStop}%
\bibitem [{\citenamefont {Navarro}\ \emph {et~al.}(1996)\citenamefont {Navarro}, \citenamefont {Frenk},\ and\ \citenamefont {White}}]{Navarro:1995iw}%
  \BibitemOpen
  \bibfield  {author} {\bibinfo {author} {\bibfnamefont {J.~F.}\ \bibnamefont {Navarro}}, \bibinfo {author} {\bibfnamefont {C.~S.}\ \bibnamefont {Frenk}}, \ and\ \bibinfo {author} {\bibfnamefont {S.~D.~M.}\ \bibnamefont {White}},\ }\href {\doibase 10.1086/177173} {\bibfield  {journal} {\bibinfo  {journal} {Astrophys. J.}\ }\textbf {\bibinfo {volume} {462}},\ \bibinfo {pages} {563} (\bibinfo {year} {1996})},\ \Eprint {http://arxiv.org/abs/astro-ph/9508025} {arXiv:astro-ph/9508025} \BibitemShut {NoStop}%
\bibitem [{\citenamefont {Prada}\ \emph {et~al.}(2012)\citenamefont {Prada}, \citenamefont {Klypin}, \citenamefont {Cuesta}, \citenamefont {Betancort-Rijo},\ and\ \citenamefont {Primack}}]{Prada:2011jf}%
  \BibitemOpen
  \bibfield  {author} {\bibinfo {author} {\bibfnamefont {F.}~\bibnamefont {Prada}}, \bibinfo {author} {\bibfnamefont {A.~A.}\ \bibnamefont {Klypin}}, \bibinfo {author} {\bibfnamefont {A.~J.}\ \bibnamefont {Cuesta}}, \bibinfo {author} {\bibfnamefont {J.~E.}\ \bibnamefont {Betancort-Rijo}}, \ and\ \bibinfo {author} {\bibfnamefont {J.}~\bibnamefont {Primack}},\ }\href {\doibase 10.1111/j.1365-2966.2012.21007.x} {\bibfield  {journal} {\bibinfo  {journal} {Mon. Not. Roy. Astron. Soc.}\ }\textbf {\bibinfo {volume} {423}},\ \bibinfo {pages} {3018} (\bibinfo {year} {2012})},\ \Eprint {http://arxiv.org/abs/1104.5130} {arXiv:1104.5130 [astro-ph.CO]} \BibitemShut {NoStop}%
\bibitem [{\citenamefont {van~den Bosch}\ \emph {et~al.}(2005)\citenamefont {van~den Bosch}, \citenamefont {Tormen},\ and\ \citenamefont {Giocoli}}]{vandenBosch:2004zs}%
  \BibitemOpen
  \bibfield  {author} {\bibinfo {author} {\bibfnamefont {F.~C.}\ \bibnamefont {van~den Bosch}}, \bibinfo {author} {\bibfnamefont {G.}~\bibnamefont {Tormen}}, \ and\ \bibinfo {author} {\bibfnamefont {C.}~\bibnamefont {Giocoli}},\ }\href {\doibase 10.1111/j.1365-2966.2005.08964.x} {\bibfield  {journal} {\bibinfo  {journal} {Mon. Not. Roy. Astron. Soc.}\ }\textbf {\bibinfo {volume} {359}},\ \bibinfo {pages} {1029} (\bibinfo {year} {2005})},\ \Eprint {http://arxiv.org/abs/astro-ph/0409201} {arXiv:astro-ph/0409201} \BibitemShut {NoStop}%
\bibitem [{\citenamefont {Hiroshima}\ \emph {et~al.}(2018)\citenamefont {Hiroshima}, \citenamefont {Ando},\ and\ \citenamefont {Ishiyama}}]{Hiroshima:2018kfv}%
  \BibitemOpen
  \bibfield  {author} {\bibinfo {author} {\bibfnamefont {N.}~\bibnamefont {Hiroshima}}, \bibinfo {author} {\bibfnamefont {S.}~\bibnamefont {Ando}}, \ and\ \bibinfo {author} {\bibfnamefont {T.}~\bibnamefont {Ishiyama}},\ }\href {\doibase 10.1103/PhysRevD.97.123002} {\bibfield  {journal} {\bibinfo  {journal} {Phys. Rev. D}\ }\textbf {\bibinfo {volume} {97}},\ \bibinfo {pages} {123002} (\bibinfo {year} {2018})},\ \Eprint {http://arxiv.org/abs/1803.07691} {arXiv:1803.07691 [astro-ph.CO]} \BibitemShut {NoStop}%
\bibitem [{\citenamefont {Ishiyama}\ \emph {et~al.}(2015)\citenamefont {Ishiyama}, \citenamefont {Enoki}, \citenamefont {Kobayashi}, \citenamefont {Makiya}, \citenamefont {Nagashima},\ and\ \citenamefont {Oogi}}]{Ishiyama:2014gla}%
  \BibitemOpen
  \bibfield  {author} {\bibinfo {author} {\bibfnamefont {T.}~\bibnamefont {Ishiyama}}, \bibinfo {author} {\bibfnamefont {M.}~\bibnamefont {Enoki}}, \bibinfo {author} {\bibfnamefont {M.~A.~R.}\ \bibnamefont {Kobayashi}}, \bibinfo {author} {\bibfnamefont {R.}~\bibnamefont {Makiya}}, \bibinfo {author} {\bibfnamefont {M.}~\bibnamefont {Nagashima}}, \ and\ \bibinfo {author} {\bibfnamefont {T.}~\bibnamefont {Oogi}},\ }\href {\doibase 10.1093/pasj/psv021} {\bibfield  {journal} {\bibinfo  {journal} {Publ. Astron. Soc. Jap.}\ }\textbf {\bibinfo {volume} {67}},\ \bibinfo {pages} {61} (\bibinfo {year} {2015})},\ \Eprint {http://arxiv.org/abs/1412.2860} {arXiv:1412.2860 [astro-ph.CO]} \BibitemShut {NoStop}%
\bibitem [{\citenamefont {Hiroshima}\ \emph {et~al.}(2022)\citenamefont {Hiroshima}, \citenamefont {Ando},\ and\ \citenamefont {Ishiyama}}]{Hiroshima:2022khy}%
  \BibitemOpen
  \bibfield  {author} {\bibinfo {author} {\bibfnamefont {N.}~\bibnamefont {Hiroshima}}, \bibinfo {author} {\bibfnamefont {S.}~\bibnamefont {Ando}}, \ and\ \bibinfo {author} {\bibfnamefont {T.}~\bibnamefont {Ishiyama}},\ }\href@noop {} {\  (\bibinfo {year} {2022})},\ \Eprint {http://arxiv.org/abs/2206.01358} {arXiv:2206.01358 [astro-ph.CO]} \BibitemShut {NoStop}%
\bibitem [{\citenamefont {Penarrubia}\ \emph {et~al.}(2010)\citenamefont {Penarrubia}, \citenamefont {Benson}, \citenamefont {Walker}, \citenamefont {Gilmore}, \citenamefont {McConnachie},\ and\ \citenamefont {Mayer}}]{Penarrubia:2010jk}%
  \BibitemOpen
  \bibfield  {author} {\bibinfo {author} {\bibfnamefont {J.}~\bibnamefont {Penarrubia}}, \bibinfo {author} {\bibfnamefont {A.~J.}\ \bibnamefont {Benson}}, \bibinfo {author} {\bibfnamefont {M.~G.}\ \bibnamefont {Walker}}, \bibinfo {author} {\bibfnamefont {G.}~\bibnamefont {Gilmore}}, \bibinfo {author} {\bibfnamefont {A.}~\bibnamefont {McConnachie}}, \ and\ \bibinfo {author} {\bibfnamefont {L.}~\bibnamefont {Mayer}},\ }\href {\doibase 10.1111/j.1365-2966.2010.16762.x} {\bibfield  {journal} {\bibinfo  {journal} {Mon. Not. Roy. Astron. Soc.}\ }\textbf {\bibinfo {volume} {406}},\ \bibinfo {pages} {1290} (\bibinfo {year} {2010})},\ \Eprint {http://arxiv.org/abs/1002.3376} {arXiv:1002.3376 [astro-ph.GA]} \BibitemShut {NoStop}%
\bibitem [{\citenamefont {Springel}\ \emph {et~al.}(2008)\citenamefont {Springel}, \citenamefont {Wang}, \citenamefont {Vogelsberger}, \citenamefont {Ludlow}, \citenamefont {Jenkins}, \citenamefont {Helmi}, \citenamefont {Navarro}, \citenamefont {Frenk},\ and\ \citenamefont {White}}]{Springel:2008cc}%
  \BibitemOpen
  \bibfield  {author} {\bibinfo {author} {\bibfnamefont {V.}~\bibnamefont {Springel}}, \bibinfo {author} {\bibfnamefont {J.}~\bibnamefont {Wang}}, \bibinfo {author} {\bibfnamefont {M.}~\bibnamefont {Vogelsberger}}, \bibinfo {author} {\bibfnamefont {A.}~\bibnamefont {Ludlow}}, \bibinfo {author} {\bibfnamefont {A.}~\bibnamefont {Jenkins}}, \bibinfo {author} {\bibfnamefont {A.}~\bibnamefont {Helmi}}, \bibinfo {author} {\bibfnamefont {J.~F.}\ \bibnamefont {Navarro}}, \bibinfo {author} {\bibfnamefont {C.~S.}\ \bibnamefont {Frenk}}, \ and\ \bibinfo {author} {\bibfnamefont {S.~D.~M.}\ \bibnamefont {White}},\ }\href {\doibase 10.1111/j.1365-2966.2008.14066.x} {\bibfield  {journal} {\bibinfo  {journal} {Mon. Not. Roy. Astron. Soc.}\ }\textbf {\bibinfo {volume} {391}},\ \bibinfo {pages} {1685} (\bibinfo {year} {2008})},\ \Eprint {http://arxiv.org/abs/0809.0898} {arXiv:0809.0898 [astro-ph]} \BibitemShut {NoStop}%
\bibitem [{\citenamefont {Bird}\ \emph {et~al.}(2011)\citenamefont {Bird}, \citenamefont {Peiris}, \citenamefont {Viel},\ and\ \citenamefont {Verde}}]{Bird_2011}%
  \BibitemOpen
  \bibfield  {author} {\bibinfo {author} {\bibfnamefont {S.}~\bibnamefont {Bird}}, \bibinfo {author} {\bibfnamefont {H.~V.}\ \bibnamefont {Peiris}}, \bibinfo {author} {\bibfnamefont {M.}~\bibnamefont {Viel}}, \ and\ \bibinfo {author} {\bibfnamefont {L.}~\bibnamefont {Verde}},\ }\href {\doibase 10.1111/j.1365-2966.2011.18245.x} {\bibfield  {journal} {\bibinfo  {journal} {Monthly Notices of the Royal Astronomical Society}\ }\textbf {\bibinfo {volume} {413}},\ \bibinfo {pages} {1717–1728} (\bibinfo {year} {2011})}\BibitemShut {NoStop}%
\bibitem [{\citenamefont {Mather}\ \emph {et~al.}(1994)\citenamefont {Mather} \emph {et~al.}}]{Mather:1993ij}%
  \BibitemOpen
  \bibfield  {author} {\bibinfo {author} {\bibfnamefont {J.~C.}\ \bibnamefont {Mather}} \emph {et~al.},\ }\href {\doibase 10.1086/173574} {\bibfield  {journal} {\bibinfo  {journal} {Astrophys. J.}\ }\textbf {\bibinfo {volume} {420}},\ \bibinfo {pages} {439} (\bibinfo {year} {1994})}\BibitemShut {NoStop}%
\bibitem [{\citenamefont {Fixsen}\ \emph {et~al.}(1996)\citenamefont {Fixsen}, \citenamefont {Cheng}, \citenamefont {Gales}, \citenamefont {Mather}, \citenamefont {Shafer},\ and\ \citenamefont {Wright}}]{Fixsen:1996nj}%
  \BibitemOpen
  \bibfield  {author} {\bibinfo {author} {\bibfnamefont {D.~J.}\ \bibnamefont {Fixsen}}, \bibinfo {author} {\bibfnamefont {E.~S.}\ \bibnamefont {Cheng}}, \bibinfo {author} {\bibfnamefont {J.~M.}\ \bibnamefont {Gales}}, \bibinfo {author} {\bibfnamefont {J.~C.}\ \bibnamefont {Mather}}, \bibinfo {author} {\bibfnamefont {R.~A.}\ \bibnamefont {Shafer}}, \ and\ \bibinfo {author} {\bibfnamefont {E.~L.}\ \bibnamefont {Wright}},\ }\href {\doibase 10.1086/178173} {\bibfield  {journal} {\bibinfo  {journal} {Astrophys. J.}\ }\textbf {\bibinfo {volume} {473}},\ \bibinfo {pages} {576} (\bibinfo {year} {1996})},\ \Eprint {http://arxiv.org/abs/astro-ph/9605054} {arXiv:astro-ph/9605054} \BibitemShut {NoStop}%
\bibitem [{\citenamefont {Nakama}\ \emph {et~al.}(2017{\natexlab{b}})\citenamefont {Nakama}, \citenamefont {Chluba},\ and\ \citenamefont {Kamionkowski}}]{Nakama:2017ohe}%
  \BibitemOpen
  \bibfield  {author} {\bibinfo {author} {\bibfnamefont {T.}~\bibnamefont {Nakama}}, \bibinfo {author} {\bibfnamefont {J.}~\bibnamefont {Chluba}}, \ and\ \bibinfo {author} {\bibfnamefont {M.}~\bibnamefont {Kamionkowski}},\ }\href {\doibase 10.1103/PhysRevD.95.121302} {\bibfield  {journal} {\bibinfo  {journal} {Phys. Rev. D}\ }\textbf {\bibinfo {volume} {95}},\ \bibinfo {pages} {121302} (\bibinfo {year} {2017}{\natexlab{b}})},\ \Eprint {http://arxiv.org/abs/1703.10559} {arXiv:1703.10559 [astro-ph.CO]} \BibitemShut {NoStop}%
\bibitem [{\citenamefont {Shipp}\ \emph {et~al.}(2023)\citenamefont {Shipp} \emph {et~al.}}]{S5FIRE:2022ylq}%
  \BibitemOpen
  \bibfield  {author} {\bibinfo {author} {\bibfnamefont {N.}~\bibnamefont {Shipp}} \emph {et~al.} (\bibinfo {collaboration} {S5 \& FIRE}),\ }\href {\doibase 10.3847/1538-4357/acc582} {\bibfield  {journal} {\bibinfo  {journal} {Astrophys. J.}\ }\textbf {\bibinfo {volume} {949}},\ \bibinfo {pages} {44} (\bibinfo {year} {2023})},\ \Eprint {http://arxiv.org/abs/2208.02255} {arXiv:2208.02255 [astro-ph.GA]} \BibitemShut {NoStop}%
\bibitem [{\citenamefont {Esteban}\ \emph {et~al.}(2024)\citenamefont {Esteban}, \citenamefont {Peter},\ and\ \citenamefont {Kim}}]{Esteban:2023xpk}%
  \BibitemOpen
  \bibfield  {author} {\bibinfo {author} {\bibfnamefont {I.}~\bibnamefont {Esteban}}, \bibinfo {author} {\bibfnamefont {A.~H.~G.}\ \bibnamefont {Peter}}, \ and\ \bibinfo {author} {\bibfnamefont {S.~Y.}\ \bibnamefont {Kim}},\ }\href {\doibase 10.1103/PhysRevD.110.123013} {\bibfield  {journal} {\bibinfo  {journal} {Phys. Rev. D}\ }\textbf {\bibinfo {volume} {110}},\ \bibinfo {pages} {123013} (\bibinfo {year} {2024})},\ \Eprint {http://arxiv.org/abs/2306.04674} {arXiv:2306.04674 [astro-ph.CO]} \BibitemShut {NoStop}%
\bibitem [{\citenamefont {Ragavendra}\ \emph {et~al.}(2024)\citenamefont {Ragavendra}, \citenamefont {Sarkar},\ and\ \citenamefont {Sethi}}]{ragavendra2024}%
  \BibitemOpen
  \bibfield  {author} {\bibinfo {author} {\bibfnamefont {H.~V.}\ \bibnamefont {Ragavendra}}, \bibinfo {author} {\bibfnamefont {A.~K.}\ \bibnamefont {Sarkar}}, \ and\ \bibinfo {author} {\bibfnamefont {S.~K.}\ \bibnamefont {Sethi}},\ }\href@noop {} {\  (\bibinfo {year} {2024})},\ \Eprint {http://arxiv.org/abs/2404.00933} {arXiv:2404.00933 [astro-ph.CO]} \BibitemShut {NoStop}%
\bibitem [{\citenamefont {Sharma}\ \emph {et~al.}(2024)\citenamefont {Sharma}, \citenamefont {Lesgourgues},\ and\ \citenamefont {Byrnes}}]{Sharma:2024img}%
  \BibitemOpen
  \bibfield  {author} {\bibinfo {author} {\bibfnamefont {D.}~\bibnamefont {Sharma}}, \bibinfo {author} {\bibfnamefont {J.}~\bibnamefont {Lesgourgues}}, \ and\ \bibinfo {author} {\bibfnamefont {C.~T.}\ \bibnamefont {Byrnes}},\ }\href {\doibase 10.1088/1475-7516/2024/07/090} {\bibfield  {journal} {\bibinfo  {journal} {JCAP}\ }\textbf {\bibinfo {volume} {07}},\ \bibinfo {pages} {090} (\bibinfo {year} {2024})},\ \Eprint {http://arxiv.org/abs/2404.18474} {arXiv:2404.18474 [astro-ph.CO]} \BibitemShut {NoStop}%
\bibitem [{\citenamefont {Byrnes}\ \emph {et~al.}(2024)\citenamefont {Byrnes}, \citenamefont {Lesgourgues},\ and\ \citenamefont {Sharma}}]{Byrnes:2024vjt}%
  \BibitemOpen
  \bibfield  {author} {\bibinfo {author} {\bibfnamefont {C.~T.}\ \bibnamefont {Byrnes}}, \bibinfo {author} {\bibfnamefont {J.}~\bibnamefont {Lesgourgues}}, \ and\ \bibinfo {author} {\bibfnamefont {D.}~\bibnamefont {Sharma}},\ }\href@noop {} {\  (\bibinfo {year} {2024})},\ \Eprint {http://arxiv.org/abs/2404.18475} {arXiv:2404.18475 [astro-ph.CO]} \BibitemShut {NoStop}%
\bibitem [{\citenamefont {Nadler}\ \emph {et~al.}(2025)\citenamefont {Nadler}, \citenamefont {Gluscevic},\ and\ \citenamefont {Benson}}]{Nadler:2025crd}%
  \BibitemOpen
  \bibfield  {author} {\bibinfo {author} {\bibfnamefont {E.~O.}\ \bibnamefont {Nadler}}, \bibinfo {author} {\bibfnamefont {V.}~\bibnamefont {Gluscevic}}, \ and\ \bibinfo {author} {\bibfnamefont {A.}~\bibnamefont {Benson}},\ }\href@noop {} {\  (\bibinfo {year} {2025})},\ \Eprint {http://arxiv.org/abs/2507.16889} {arXiv:2507.16889 [astro-ph.CO]} \BibitemShut {NoStop}%
\end{thebibliography}%
%%%%%%%%%%%%%%%%%%%%%%%%%%%%%%%%%%%%%%%%%%%%%%%%%%%%%%%%%%%%%%%%%%%%%%%%%%%%%%%

%%%%%%%%%%%%%%%%%%%%%%%%%%%%%%%%%%%%%%%%%%%%%%%%%%%%%%%%%%%%%%%%%%%%%%%%%%%%%%%%
\end{document}